\documentclass[lettersize,journal]{IEEEtran}
\usepackage{amsmath,amsfonts,amsthm}
\usepackage{algorithmicx}
\usepackage{algorithm}
\usepackage{array}
\usepackage[caption=false,font=normalsize,labelfont=sf,textfont=sf]{subfig}
\usepackage{textcomp}
\usepackage{stfloats}
\usepackage{url}
\usepackage{verbatim}
\usepackage{graphicx}
\usepackage{cite}
\usepackage{multirow}
\usepackage{enumitem}
\usepackage{makecell}
\usepackage{setspace}
\usepackage[noend]{algpseudocode}
\usepackage{amssymb}
\usepackage{color}

\begin{document}

\newtheorem{theorem}{Theorem}

\newtheorem{definition}{Definition}
\newtheorem{lemma}{Lemma}
\newtheorem{case}{Case}
\setlist{leftmargin=6mm}
\setlength{\itemsep}{0.4ex}

\title{FedTracker: Furnishing Ownership Verification and Traceability for Federated Learning Model}

\author{Shuo~Shao, Wenyuan~Yang, Hanlin~Gu, Zhan~Qin, Lixin~Fan, Qiang~Yang,~\IEEEmembership{Fellow,~IEEE,} \\Kui~Ren,~\IEEEmembership{Fellow,~IEEE}%
\thanks{S. Shao, Z. Qin, K. Ren are with the State Key Laboratory of Blockchain and Data Security, Zhejiang University, China, Email address: {\tt \{shaoshuo\_ss, qinzhan, kuiren\}@zju.edu.cn}}%
\thanks{W. Yang is with the School of Cyber Science and Technology, Sun Yat-sen University, China, Email address: yangwy56@mail.sysu.edu.cn}%
\thanks{H. Gu and L. Fan are with the Webank AI Lab, China, Email address: ghltsl123@gmail.com, lixinfan@webank.com}%
\thanks{Q. Yang is with the Computer Science and Engineering Department, Hong Kong University of Science and Technology, Email address: qyang@cse.ust.hk}
\thanks{The first two authors have equal contributions.}
\thanks{Zhan Qin is the corresponding author.}
}

\markboth{IEEE Transactions on Dependable and Secure Computing}%
{Shao \MakeLowercase{\textit{et al.}}: FedTracker: Furnishing Ownership Verification and Traceability for Federated Learning Model}


\maketitle

\begin{abstract}
Federated learning (FL) is a distributed machine learning paradigm allowing multiple clients to collaboratively train a global model without sharing their local data. However, FL entails exposing the model to various participants. This poses a risk of unauthorized model distribution or resale by the malicious client, compromising the intellectual property rights of the FL group. To deter such misbehavior, it is essential to establish a mechanism for verifying the ownership of the model and as well tracing its origin to the leaker among the FL participants. In this paper, we present FedTracker, the first FL model protection framework that provides both ownership verification and traceability. FedTracker adopts a bi-level protection scheme consisting of global watermark mechanism and local fingerprint mechanism. The former authenticates the ownership of the global model, while the latter identifies which client the model is derived from. FedTracker leverages Continual Learning (CL) principles to embed the watermark in a way that preserves the utility of the FL model on both primitive task and watermark task. FedTracker also devises a novel metric to better discriminate different fingerprints. Experimental results show FedTracker is effective in ownership verification, traceability, and maintains good fidelity and robustness against various watermark removal attacks.
\end{abstract}




\section{Introduction}
\label{sec:intro}

\IEEEPARstart{F}{ederated} Learning (FL)~\cite{mcmahan2017communication} has turned into one of the most popular frameworks for distributedly training Deep Learning (DL) models without sharing private messages. FL is nowadays widely applied in various privacy-sensitive domains, such as object detection~\cite{liu2020fedvision, yu2019federated}, medical image analysis~\cite{liu2021feddg,kumar2021blockchain, Xu_2022_CVPR}, recommendation systems~\cite{wahab2022federated,yang2020federated} and so on. FL meets the privacy needs of the data owners while producing high-quality models for the community. However, collaborative training in FL makes the model publicly available to multiple participants, which may lead to the model leakage issue in FL.

Model leakage in FL refers to the malicious distribution or unauthorized sale of FL models by the individual Client for personal gain. These malicious Clients are called the model leakers. Such misbehavior infringes the legitimate copyright of the entire FL group, including innocent Clients and Servers. Detecting copyright infringement and protecting intellectual property rights has become a critical issue in FL. There are two major concerns in FL copyright protection, ownership verification and traceability. Ownership verification involves verifying the ownership of the FL model when it is illegally distributed and charging the unauthorized user outside the FL group with tort. The FL model should be considered as the intellectual property of the FL group instead of any individual. Any legal use of the FL model should be made with the consent of the FL group. Traceability refers to disclosing the identity of the culprit among the FL participants who illegally distribute the model to third parties outside the FL group. It is a new requirement for protecting the copyright of the FL model, which to our best knowledge has not been addressed before.

To protect the copyright of Deep Neural Networks (DNN), the DNN watermarks~\cite{xueIntellectualPropertyProtection2021, sun2023deep}
, including \emph{parameter-based} watermark and \emph{backdoor-based} watermark, are proposed. \emph{Parameter-based} watermark methods embed personal watermarks into the parameters or the distribution of the parameters of the DNN models~\cite{uchida2017embedding,darvish2019deepsigns,fan2019rethinking,li2022encryption}. It can provide multi-bit information for verification, but the extraction of \emph{parameter-based} watermarks need white-box access to the model, which limits the application. On the contrary, \emph{backdoor-based} watermark methods~\cite{adi2018turning, bansalCertifiedNeuralNetwork2022,li2019how,laoIdentificationDeepNeural2022} take advantage of backdoor attack~\cite{gu2017badnets} to insert specific triggers to identify the ownership of the model. Although \emph{backdoor-based} watermark can be verified through black-box access, it is usually zero-bit. This means the watermark can only demonstrate the presence or absence of the watermark instead of a bit string representing identity.

\begin{figure*}[t!]
  \centering
  \includegraphics[width=0.92\linewidth]{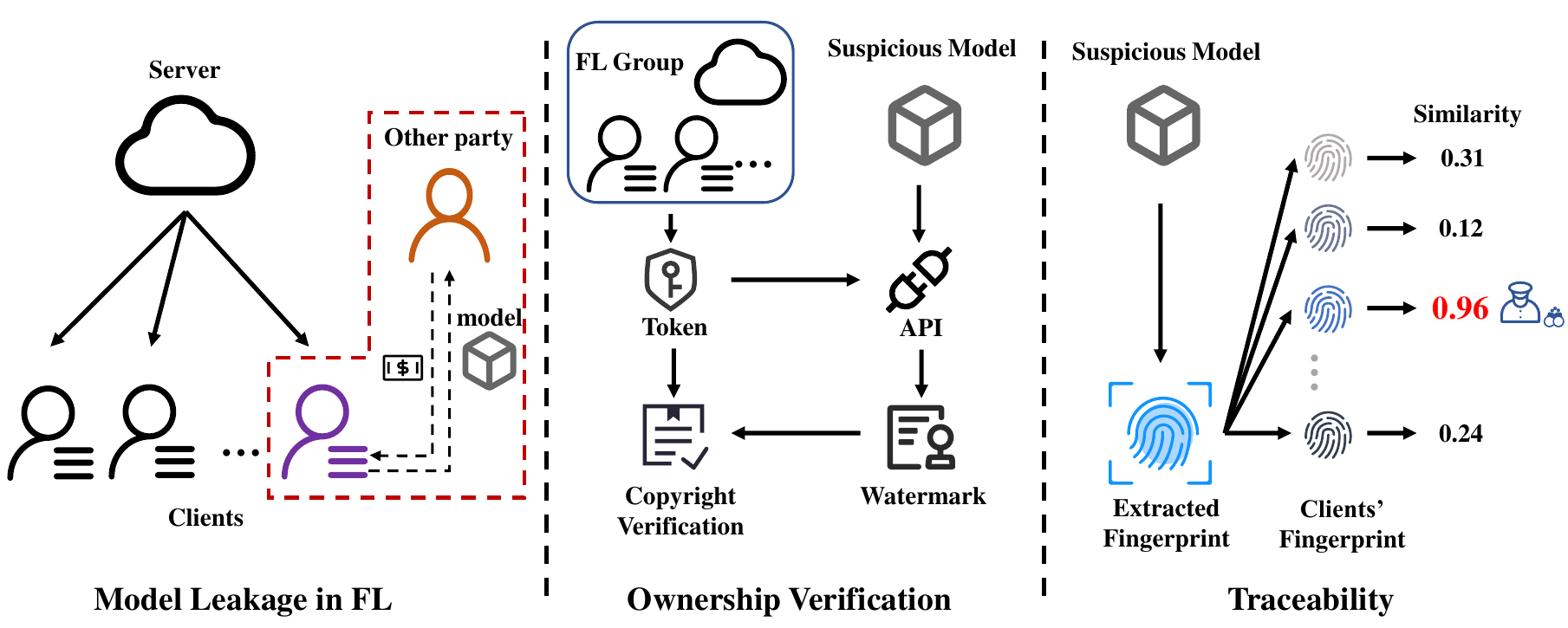}
  \caption{Illustration of model leakage in FL and the solution framework. \textbf{Left}: the malicious Client can sell the model to other parties. \textbf{Middle}: The FL group needs to verify the ownership through API access. \textbf{Right}: The fingerprint is extracted from the model and compared with all the fingerprints of Clients to trace the leaker.}
  \vspace{-15pt}
  \label{fig:problem}
\end{figure*}

Some recent works also explore applying DNN watermarks to FL copyright protection. WAFFLE~\cite{tekgul2021waffle} is the first watermarking scheme designed for FL, in which the Server is responsible for embedding the \emph{backdoor-based} watermarks into the FL model. WAFFLE proposes to conduct a retraining step after federated aggregation to embed the watermark. Liu et al.~\cite{liuSecureFederatedLearning2021, yang2023watermarking} and Li et al.~\cite{liFedIPROwnershipVerification2022} propose to entitle the Clients to embed both \emph{parameter-based} and \emph{backdoor-based} watermarks into the global model. However, existing works only focus on ownership verification. None of them can provide traceability of the model leakers for FL models. Even if the FL group can prove the ownership of the FL model in the event of infringement, they are not capable of detecting the model leakers hidden among them.

We aim to design an FL copyright protection framework that can furnish both ownership verification and traceability for the FL model. In this paper, we present FedTracker, the first FL model protection framework that offers both. The illustration of the problem and our framework is shown in Fig. \ref{fig:problem}. Specifically, we design a bi-level protection scheme that enables ownership verification and traceability respectively. On one level, FedTracker embeds a global watermark into the aggregated global model in each iteration. While the copyright infringement occurs, the FL group can extract the watermark and verify the ownership. On the other level, before distributing the model in each iteration, the server inserts a unique local fingerprint for each Client to identify each model. The model leaker can be traced by disclosing the identification inside the model.

However, there are two main challenges associated with implementing such a framework:
\begin{itemize}
  \item It is challenging to preserve the utility of the model in the condition of a lack of natural data. Since the server is assumed to have no access to the training data, the server needs to embed the watermark trigger set which is out of distribution of natural data. This may significantly compromise the model's utility.
  \item It is challenging to guarantee the discrimination of different models. Since we need to embed and compare multiple different fingerprints from different Clients, it puts a higher requirement on the identifiability and robustness of the Clients' fingerprints.
\end{itemize}

To tackle the challenges mentioned above, first, we propose a watermark embedding algorithm based on Continual Learning (CL)~\cite{kirkpatrick2017overcoming} to improve the utility of the watermarked model. We regard the primitive task of FL and the watermark embedding task as learning data from different domains. The utility drop can be attributed to catastrophic forgetting~\cite{kirkpatrick2017overcoming} while learning a different task. Inspired by the Gradient Episodic Memory (GEM)~\cite{lopez-pazGradientEpisodicMemory2017}, we design a CL-based embedding algorithm for the FL scenario and thereby, the utility can be better preserved during watermark embedding. We introduce our CL-based watermark embedding algorithm in Section \ref{sec:embed}.

Second, to enhance the differentiation between different models of different Clients, we conduct the fingerprint generation by minimizing the maximal distance between fingerprints. We also propose a Fingerprint Similarity Score (FSS) to measure the similarity of the extracted fingerprint and Clients' fingerprints. Compared with traditional hamming distance whose output is discrete, FSS can provide more fine-grained measurement because the output of FSS is continuous. Thus, using FSS can better discriminate different fingerprints. Moreover, we embed the watermark and fingerprints respectively into different parts of the model to lower the mutual impact and improve the effectiveness of both mechanisms. The details are present in Section \ref{sec:distinguish}.

\subsection{Contributions}

Our main contributions can be summarized as follows:
\begin{itemize}
  \item We propose the first FL ownership verification and traceability framework called FedTracker. In FedTracker, the bi-level protection scheme, including the global watermark mechanism and the local fingerprint mechanism, can provide ownership verification and traceability respectively. Experimental results show that FedTracker can also provide tremendous fidelity and robustness against various attacks.
  \item Inspired by the motivation of Continual Learning (CL), we design a CL-based watermark embedding algorithm for the FL scenario. It can maintain the utility of the FL model on both the primitive task and the watermark task while embedding the watermark.
  \item We propose a fingerprint generation method based on optimization and a novel metric named Fingerprint Similarity Score (FSS). The two methods can help better distinguish different fingerprints and improve the effectiveness of tracing the culprit. 
\end{itemize}

\subsection{Roadmap}

The rest of this paper is organized as follows:  Section \ref{sec:back} provides an introduction to the background knowledge of FedTracker. Section \ref{sec:problem} gives the formulation of the copyright protection problem in FL, including the definition of ownership verification and traceability and the threat model. Section \ref{sec:design} proposes the procedure of FedTracker, and the solutions to the key challenges proposed above. For a detailed implementation description of FedTracker, please refer to Section \ref{sec:implementation}. Subsequently, in Section \ref{sec:experiment}, we present the empirical experiments conducted on FedTracker's performance evaluation, and Section \ref{sec:conclusion} concludes this paper.

\section{Background}
\label{sec:back}                                                                              

\subsection{Federated Learning}

Federated Learning (FL)~\cite{mcmahan2017communication} is a distributed machine learning paradigm that trains a Deep Learning model across multiple participants holding local data samples. In FL, intermediate results such as gradients rather than sensitive data are shared between data owners. This approach can protect the privacy and security of the data owner while enabling collaborative learning among different parties.

Client-Server FL is the most popular FL paradigm in practice. 
In the Client-Server FL~\cite{yang2019federatedlearning, fereidooni2021safelearn}, there are two different parties involved: Client and Server. Clients $C=\{c_i\}_{i=1}^K$ are the data owners in FL who hold a bunch of personal data. $K$ is the number of Clients. The server is responsible for collecting the intermediate local gradients $L=\{g_i\}_{i=1}^K$ from Clients and aggregating them into a global model $M_g$. 

In each iteration $t$, Clients and Server do
\begin{enumerate}
  \item The Clients get the global model $M^{t-1}_g$ of the $t-1$ iteration from the Server. We use the superscript $t-1$ to mark the iteration number and the subscript represents the number of Clients or the global model (the subscript is denoted as $g$).
  \item Each Client  $c_i$ copy the global model onto its local model $M_i^{t-1} \leftarrow M_g^{t-1}$, and then utilizes its private dataset $D_i$ to train the local model $M^{t-1}_i$ for several epochs. For a given loss function $\mathcal{L}(\cdot)$, the local gradients $g_i^t$ can be calculated using the following equation.
  \begin{equation}
    g_i^t = \frac{\partial\mathcal{L}(M_i^{t-1}, D_i)}{\partial M_i^{t-1}}.
  \end{equation}
  Then, the Clients apply gradient descent to update the model to $M_i^t$.
  \begin{equation}
      M_i^t = M_i^{t-1} - \lambda_i g_i^t.
  \end{equation}
  where $\lambda_i$ is the local learning rate of the Client $i$. After that, the Client sends the local model $M_i^t$ of $t$-th iteration to the Server for aggregation.
  \item Server collects the local models and performs the federated aggregation algorithm to aggregate the local gradients into a new global model $M^t_g$. FedAvg~\cite{mcmahan2017communication} is a widely used aggregation algorithm in FL. In particular, FedAvg utilizes Equation (\ref{eq:fedavg}) to aggregate the local models.
  \begin{equation}
    \label{eq:fedavg}
    M_g^{t} = \sum_{i=1}^{K}\frac{|D_i|}{\sum_{j=1}^{K}|D_j|}M_i^{t}.
  \end{equation}
  \item Server sends the global model $M^t_g$ to all the Clients. Clients acquire the new global model and begin the next iteration of FL training.
\end{enumerate} 

\subsection{DNN Watermarking Methods}

Broadly speaking, existing DNN watermarking methods can be categorized as \emph{parameter-based} watermarks and \emph{backdoor-based} watermarks~\cite{regazzoni2021protecting, xueIntellectualPropertyProtection2021}.

\vspace{0.3em}
\noindent\textbf{Parameter-based Watermarks}~\cite{uchida2017embedding,darvish2019deepsigns,fan2019rethinking,li2022encryption}. \emph{Parameter-based} watermarks decide to embed the watermarks, which can be represented as an $N$-bit string $F \in \{0,1\}^N$, directly into the parameters $W$ of the models. The embedding of the watermark can be considered a binary classification problem and the watermark can be fitted by adding a specific regularization term into the loss function.

During the verification, the watermark $F^{'}$ is extracted from the suspicious model. $F^{'}$ is then compared with the designated watermarks $F$. If the distance $D(F, F^{'})$ (usually hamming distance) of the two watermarks is less than a threshold $\varepsilon_P$, the ownership of the model can be verified. 

\emph{Parameter-based} watermarks can provide multi-bit information, but the extraction of watermarks needs white-box access to the suspicious model.

\vspace{0.3em}
\noindent\textbf{Backdoor-based Watermarks}~\cite{adi2018turning, bansalCertifiedNeuralNetwork2022,li2019how,laoIdentificationDeepNeural2022}. \emph{Backdoor-based} watermarks utilize the backdoor attack~\cite{li2022backdoor} to insert specific trigger set $D_T$ into the DNN model. The backdoor attack leads to misclassification when encountering samples in the trigger set. The trigger set is unique to the backdoored model, thus the owner can verify its ownership by triggering the misclassification. \emph{Backdoor-based} watermarks have been applied to protect the copyright of various DL models, such as image classification models~\cite{adi2018turning}, text generation models~\cite{li2023plmmark}, and prompt~\cite{yao2024prompt}.

\emph{Backdoor-based} watermarks can be verified through black-box access, but they are typically zero-bit. This means they only indicate the presence or absence of the watermark, not a bit string that identifies the owner.

\subsection{FL Watermarking Methods}
\label{sec:flwatermark}

DNN watermarking techniques can also be employed for safeguarding the copyright of FL models~\cite{lansari2023federated}. The FL watermarking methods can be typically categorized as \emph{server-side} and \emph{Client-side} watermarking methods. 

\vspace{0.3em}
\noindent\textbf{Server-side Watermarking Methods.} WAFFLE~\cite{tekgul2021waffle} proposes the first server-side FL watermarking method. WAFFLE adds a retraining step to embed \emph{backdoor-based} watermarks into the FL global model. Besides, because the Server is assumed to lack training data, WAFFLE designs a data-free trigger set construction algorithm called WafflePattern. However, WAFFLE only focuses on ownership verification and ignores traceability. The fidelity of applying WAFFLE is also not guaranteed.

\vspace{0.3em}
\noindent\textbf{Client-side Watermarking Methods.} As for Client-side watermarking methods, Liu et al.~\cite{liuSecureFederatedLearning2021, yang2023watermarking} and Li et al.~\cite{liFedIPROwnershipVerification2022} propose to entitle the Clients to watermark the FL model. Liu et al.~\cite{liuSecureFederatedLearning2021} suggest selecting a representative Client to embed the watermark through a gradient-enhanced embedding algorithm. Li et al.~\cite{liFedIPROwnershipVerification2022} think every Client should have the ability to verify its copyright. Thus Li et al. propose FedIPR which entitles every Client to embed its own watermark. Nevertheless, what the Clients embed can hardly be monitored. The Clients can embed both the watermarks and harmful things like backdoor~\cite{bagdasaryan2020backdoor, rieger2022deepsight} into the model. Hence, the assurance of security in implementing Client-side watermarking cannot be guaranteed.

\subsection{Continual Learning}
\label{sec:cl}

Continual Learning (CL)~\cite{de2021continual}, also called life-long learning~\cite{parisi2019continual} or incremental learning~\cite{masana2022class}, is a learning technique that aims to enable a machine learning model to learn from a stream of data without forgetting the previous knowledge. CL poses several challenges for both theory and practice, such as how to measure and mitigate forgetting, how to transfer and generalize knowledge across tasks, and how to balance the stability and plasticity of the learning system.

A common approach of CL is to use the regularization techniques that constrain the model parameters to prevent catastrophic forgetting, such as elastic weight consolidation (EWC)~\cite{kirkpatrick2017overcoming}, synaptic intelligence (SI)~\cite{zenke2017continual}, or learning without forgetting (LwF)~\cite{li2017learning}. Another approach is to use replay techniques that augment the current data with previous tasks. The most notable method is class incremental learner iCaRL~\cite{rebuffi2017icarl}. iCaRL stores a subset of exemplars per class, selected to best approximate the class means in the learned feature space. Gradient episodic memory (GEM)~\cite{lopez-pazGradientEpisodicMemory2017} proposes to project the gradient direction of the current task on the feasible region outlined by previous tasks. Thus, the new task updates do not interfere with previous tasks.

\section{Problem Formulation}
\label{sec:problem}

\subsection{Definitions of Ownership Verification and Traceability}

Federated Learning enables multiple data owners to collaboratively train a DNN model in a privacy-preserving way due to the increasing concern about personal privacy security. However, FL faces the challenge of model leakage. Model leakage refers to the malicious participants in FL can stealthily copy and sell the FL model to other parties. Such misbehavior infringes the legitimate copyright of the FL group, and we should design a mechanism to verify the ownership and disclose the identity of the model leaker among the participants.

There are two major concerns in the copyright protection of the FL model, namely ownership verification and traceability. In this subsection, we give the formal definitions of them. Note that in this paper, we focus on ownership verification and traceability in model copyright protection after training the model, and the definition of traceability is different from other works trying to find abnormal Clients during training~\cite{chen2023privacy, ren2022privacy}.

\begin{definition}[Ownership verification]
Ownership verification refers to proving that the suspicious model belongs to the FL group and then charging the unauthorized user outside the FL group with tort. The suspicious model is defined as the model that is suspected to be a copy of the FL models.  When the infringement occurs, the FL group can apply the verification algorithm ${\tt Verify}(\cdot)$ to verify their ownership of the model. The ownership verification mechanism ${\tt Verify}(\tilde{M}, D_T)$ should output:
\begin{equation}
{\tt Verify}(\tilde{M}, D_T)=
\left\{
  \begin{aligned}
    &{\tt True},\tilde{M}\in \bigcup_{i=1}^K{\tt att}(M_i)\\
    &{\tt False},{\tt if\enspace otherwise}
  \end{aligned}
\right.,
\end{equation}
where $\tilde{M}$ is the suspicious model, $D_T$ is the verification token offered by the FL group, and ${\tt att}(\cdot)$ represents various attack methods which will be introduced in Section \ref{sec:threat}.
\end{definition}

\begin{definition}[Traceability]
  Traceability refers to tracing the stolen model to the malicious Client in FL. After verifying the ownership of the suspicious model, the FL group should further adopt the tracing mechanism 
  ${\tt Trace}(\cdot)$ to disclose the model leaker. Given a suspicious model $\tilde{M}$ and fingerprints of Clients $\{F_i\}_{i=1}^K$, the traceability mechanism ${\tt Trace}(\tilde{M}, \{F_i\}_{i=1}^K)$ should output the identity ($i.e.$ index) of the model leaker in FL.
  \begin{equation}
    {\tt Trace}(\tilde{M}, \{F_i\}_{i=1}^K)=j, \tilde{M}\in {\tt att}(M_j).
  \end{equation}
\end{definition}

\subsection{Threat Model}
\label{sec:threat}

Server and Clients are the two different parties in FL. Following the prior works~\cite{tekgul2021waffle}, we assume that the Server is a trusted and honest party. Some of the Clients might be malicious, while the other Clients are assumed to be honest. Malicious Clients can copy, distribute, and sell the jointly-trained model stealthily.

\vspace{0.3em}
\noindent\textbf{Adversary's assumptions:} the adversary intends to acquire a high-performance DNN model through FL, and secretly sell the model to other parties. Moreover, the adversary can try to remove the watermarks in the model. We assume that the adversary has the following capabilities:
\begin{itemize}
  \item The adversary has access to its own dataset $D_{adv}$ and the local model $M_{adv}^t$ in any iteration round $t$ of FL training. Also, the adversary can save and copy the model $M_{adv}^t$ at any time of training.
  \item The adversary follows the training procedure to maximize the utility and value of the FL model.
  \item The adversary can conduct several removal techniques to get rid of the watermarks in the DNN model. The watermark removal techniques are listed below.
\end{itemize}

\vspace{0.3em}
\noindent\textbf{Adversary's Attack Techniques:} The adversary may adopt various techniques to remove the watermarks so that it can evade the verification and tracing. Drawing on the literature on watermark attack methods proposed in preliminary works, we classify the potential techniques of adversaries into several distinct types.

\textbf{Fine-tuning}. The adversary may adopt Fine-tuning to obfuscate the model parameters and behaviors. Fine-tuning attacks refer to resuming the training procedure and modifying the model parameters using the dataset of the adversary and a low learning rate. Due to the non-convexity of deep learning~\cite{choromanska2015loss}, fine-tuning might make the deep learning model get to a different local optimum, and thus remove the watermark. Fine-tuning should be considered as a threat to the watermarks.

\textbf{Model Compression}. Model compression methods aim to eliminate the redundancy in the model parameters, and thus, the embedded watermark may also be erased during this process. In this paper, we consider two different categories of model compression methods, \emph{model pruning}~\cite{han2015learning} and \emph{quantization}. Model pruning eliminates the parameters that have little or no impact on the network performance. Model quantization reduces the numerical precision of the parameters, such as using 8-bit integers instead of 32-bit floats, to save memory bandwidth and computational cost. Both techniques could have a negative impact on the watermark with minimum accuracy loss.

\textbf{Overwriting}. Overwriting aims to remove or alter the watermark embedded in a neural network model by adding a new watermark to the model. The overwriting attack is a type of active attack. The adversary randomly generates his own watermark and uses the same method to embed the watermark into the model. The watermark in the model may be significantly affected by overwriting.

\textbf{Backdoor Mitigation}. Backdoor mitigation method~\cite{wang2019neural} is the process of detecting and removing backdoor triggers from a Deep Learning model. Since current black-box watermarking methods mostly utilize the backdoor to embed and verify the watermark, the backdoor mitigation method can be a threat to the watermark embedded in the model.

\subsection{Defender Assumptions and Objectives}
\label{sec:defender}

In our scenario, the Server in FL is the defender. The defender tries to furnish ownership verification and traceability for FL models. The capabilities of the defender are as follows.
\begin{itemize}
    \item The defender can only get the aggregated gradients or model in each iteration for the sake of preserving the data privacy. The local gradients of the Clients are noised or encrypted~\cite{bonawitz2017practical}.
    \item The defender has no access to any natural data related to the primitive DL task.
\end{itemize}

The defender needs to achieve the following four objectives, including ownership verification, traceability, fidelity, and robustness. The objectives are listed below.
\begin{itemize}
  \item \textbf{Ownership verification:} The defender can devise a mechanism to effectively authenticate the FL group's ownership on the FL model. 
  \item \textbf{Traceability:} Given an FL model, the defender can trace the holder of the model inside the FL group and catch the culprit. The defender needs to trace any model back to the leaker despite the number of malicious Clients.
  \item \textbf{Fidelity:} Fidelity refers to that the protection mechanism should have a negligible impact on the utility of the model. In other words, the accuracy of the model on the primitive task should be maintained.
  \item \textbf{Robustness:} Robustness refers to that the protection mechanism cannot be removed by the adversary. The protection mechanism should resist various attacks, such as fine-tuning, model compression, overwriting, and backdoor mitigation.
\end{itemize}

\section{Ownership Verification and Traceability in Federated Learning}
\label{sec:design}

\subsection{Overview of FedTracker}
\label{sec:overview}

In this paper, we propose FedTracker, the first FL copyright protection framework that can provide both ownership verification and traceability. FedTracker conducts the protecting mechanism from the Server-side due to the security issues of Client-side watermarking discussed in Section \ref{sec:flwatermark}. In FedTracker, a bi-level protecting mechanism is included. For the first level, a global watermark is embedded into the global model so that the models in FL can contain the global watermark for ownership verification. We utilize the backdoor-based watermark as the global watermark since the backdoor-based watermark can be verified through API access. For the second level, a unique local fingerprint is inserted into the model of each Client for further traceability. Since traceability requires distinguishing different Clients' models, the multi-bit parameter-based watermark is feasible for the local fingerprints. 

\begin{figure}[t]
  \centering
   \includegraphics[width=0.95\linewidth]{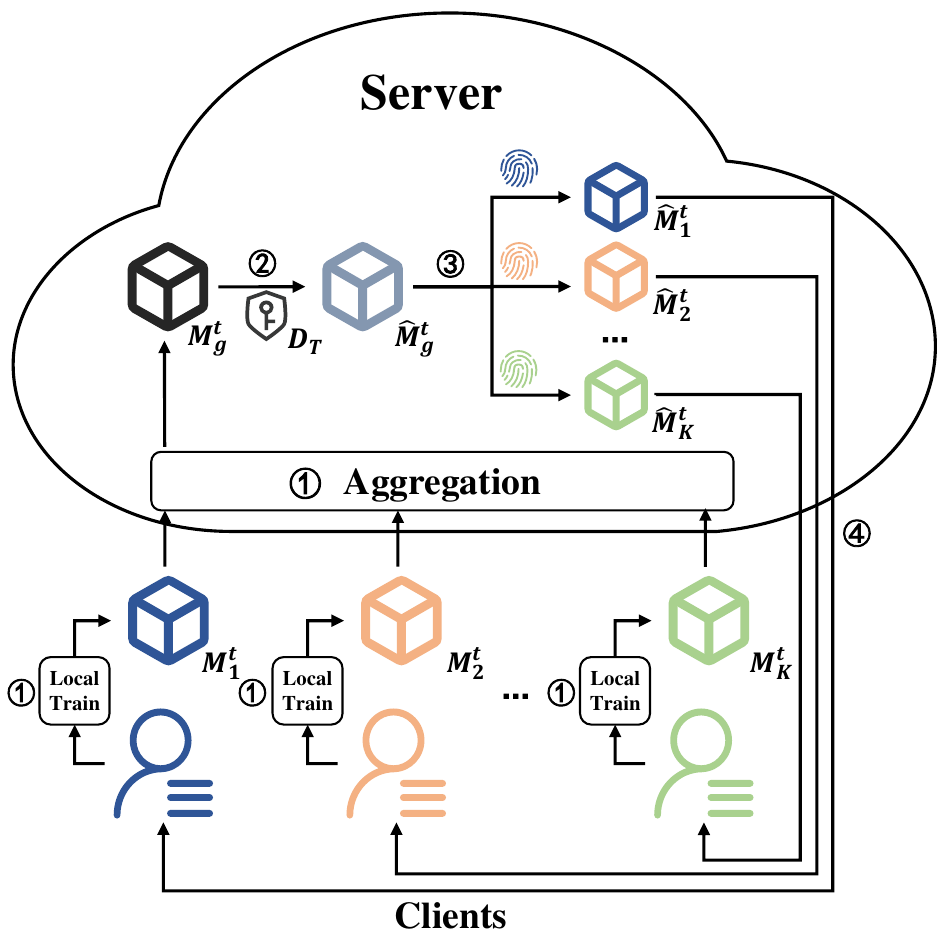}
   \vspace{-5pt}
   \caption{Workflow of FedTracker in each iteration. FedTracker consists of four stages. \normalsize{\textcircled{\scriptsize{1}}}\enspace \footnotesize{Local training and aggregation.} \normalsize{\textcircled{\scriptsize{2}}}\enspace \footnotesize{Global watermark embedding.} \normalsize{\textcircled{\scriptsize{3}}}\enspace \footnotesize{Local fingerprints insertion.} \normalsize{\textcircled{\scriptsize{4}}}\enspace \footnotesize{Model distribution.}}
   \vspace{-10pt}
   \label{fig:structure}
\end{figure}

The main procedure of FedTracker is demonstrated in Fig.~\ref{fig:structure}. After initializing the model parameters and generating the watermark and fingerprints, in each training iteration, FedTracker consists of four stages, including local training and aggregation, global watermark embedding, local fingerprint insertion, and model distribution.
\begin{itemize}
  \item \textbf{Stage 1. }(\emph{Local training and aggregation}) In this stage, each Client utilizes their own dataset to train their model locally. The Server then collects the local gradients and aggregates them into a global model via the federated aggregation algorithm, $e.g.$, FedAvg~\cite{mcmahan2017communication}.
  \item \textbf{Stage 2. }(\emph{Global watermark embedding}) In this stage, the Server embeds the global watermark into the new aggregated global model. The implementation of the global watermark mechanism is discussed in Section \ref{sec:gw}.
  \item \textbf{Stage 3. }(\emph{Local fingerprints insertion}) In this stage, the Server first generates $K$ copies of the global model ($K$ is the number of Clients). Then, a unique local fingerprint that represents a Client is inserted into each copy. The implementation of the local fingerprints mechanism is discussed in Section \ref{sec:lf}.
  \item \textbf{Stage 4. }(\emph{Model distribution}) In this stage, the Server distributes the fingerprinted models to the Clients respectively. After receiving the models, the Clients begin the next training iteration of FL.
\end{itemize}

The pseudocode of FedTracker is presented in Algorithm~\ref{algo:procedure}. As discussed in Section \ref{sec:intro}, the challenges of applying such a framework are two-fold. First, how to preserve the utility of the model while embedding a watermark without natural data? Second, how to efficiently discriminate different models of different Clients? We introduce our motivation and solutions to these challenges in Section \ref{sec:embed} and Section \ref{sec:distinguish}.

\begin{figure}[t]
\vspace{-8pt}
\begin{algorithm}[H]
  \caption{Main procedure of FedTracker}
  \label{algo:procedure}
  \hspace*{0.05in}{\bf Input:} Clients datasets $\{D_i\}_{i=1}^K$ \\
  \hspace*{0.05in}{\bf Output:} the protected Clients' models $\{\hat{M}^T_i\}_{i=1}^K$, the protected global model $\hat{M}^T_g$.
  \begin{algorithmic}[1]
  \State $D_T, \{F_i\}_{i=1}^K = {\tt Gen}()$ 
  \State $M^0_g = {\tt Initialize}()$
  \For{$i=1$ to $K$}
    \State $M^0_i = {\tt Copy}(M^0_g)$
  \EndFor
  \For{$t=1$ to $T$}
    \State // \emph{\textcircled{\footnotesize{1}} Local training and aggregation}
    \For{$i=1$ to $K$}
      \State $M^t_i = {\tt LocalTrain}(M^{t-1}_i, D_i)$
      \State $l_i^t = M^{t-1}_i - M^t_i$
    \EndFor
    \State $M_g^{t} = \sum_{i=1}^{K}\frac{|D_i|}{\sum_{j=1}^{K}|D_j|}M_i^{t}$
    \State // \emph{\textcircled{\footnotesize{2}} Global watermark embedding}
    \State $\hat{M}_g^t = {\tt GEmbed}(M_g^t, D_T)$
    \State // \emph{\textcircled{\footnotesize{3}} Local fingerprints insertion}
    \For{$i=1$ to $K$}
      \State $\hat{M}_i^t = {\tt Copy}(\hat{M}^t_g)$
      \State $\hat{M}_i^t = {\tt LInsert}(\hat{M}_i^t, F_i)$
    \EndFor
    \State // \emph{\textcircled{\footnotesize{4}} Model distribution}
    \For{$i=1$ to $K$}
      \State $M_i^t = \hat{M}_i^t$
    \EndFor
  \EndFor
  \State \Return $\{\hat{M}^T_i\}_{i=1}^K, \hat{M}^T_g$ 
\end{algorithmic}
\end{algorithm}
\vspace{-20pt}
\end{figure}

\subsection{Embedding Watermark without Natural Data}
\label{sec:embed}

\noindent\textbf{Observations and Insight}. In FedTracker, we propose to embed the global watermark from the Server side. We take advantage of \emph{backdoor-based} watermark and try to embed a special trigger set as the global watermark. The zero-bit backdoor-based watermark is feasible for ownership verification and can be verified through black-box access.

However, retraining the model on the trigger set without data related to the primitive task compromises the utility of the model, since the Server has no access to the training data. The situation of model utility degradation is more obvious when using models with Batch Normalization (BN) layers. Take WAFFLE as an example. When directly applying WAFFLE to models containing BN layers, we observe a significant utility degradation as shown in Table \ref{table:waffle}. We train the variants with BN layers of two popular convolutional neural networks (CNN), AlexNet~\cite{krizhevsky2017imagenet} and VGG-16~\cite{simonyan2014very}, on CIFAR-10~\cite{krizhevsky2009learning}. As Table \ref{table:waffle} demonstrates, the accuracies drop more than $50\%$ when applying WAFFLE to the models with BN layers, which indicates that the models are completely useless after the watermark is embedded via WAFFLE.

This can be attributed to the following two reasons. Firstly, in the initialization phase, the Server needs to conduct a data-free trigger set construction algorithm~\cite{yang2023watermarking} to generate the watermark trigger set. The data-free trigger set is out of distribution (OOD) of the training data. Retraining the model on the OOD trigger set leads to catastrophic forgetting~\cite{kirkpatrick2017overcoming} of the previous task and has a large impact on the functionality of the model. Secondly, the principle of BN layers amplifies the negative effect. BN works by applying a linear transformation to the outputs of the previous layer, such that they have zero mean and unit variance. However, the mean and variance of the trigger set are far different from natural data. Thus, the utility of the model is significantly compromised during watermark embedding.

\begin{table}[t]
  \caption{Testing accuracy (\%) of the models with BN layers when applying WAFFLE or not embedding the watermark.}
  \label{table:waffle}
  \vspace{-5pt}
  \centering
  \setstretch{1.1}
  \small
  \begin{tabular}{cc|cccc}
  \hline
  \hline
  & & \multicolumn{4}{c}{Number of Clients} \\
  & Model & 10 & 20 & 30 & 50\\
  \hline
  \multirow[c]{2}{*}{AlexNet} & No WM & 86.79 & 86.41 & 86.12 & 86.05\\
  & WAFFLE & 27.91 & 29.84 & 32.60 & 30.70\\
  \multirow[c]{2}{*}{VGG-16} & NO WM & 89.92 & 89.68 & 89.39 & 89.04\\
  & WAFFLE & 21.34 & 24.46 & 24.12 & 25.66\\
  \hline
  \hline
  \end{tabular}
  \vspace{-15pt}
\end{table}

Based on the above observations, we consider the primitive task and the watermark embedding task as two different tasks from different domains, which is exactly the motivation of Continual Learning (CL)~\cite{de2021continual}. CL aims to reduce the catastrophic forgetting of the previous task when learning a new task. Applying CL can help to reduce the negative impact on the primitive task and maintain the utility of the model on both two tasks.

\vspace{0.3em}
\noindent\textbf{CL-based Watermark Embedding Mechanism.} In FedTracker, we propose a CL-based watermark embedding algorithm. We get the inspiration from Gradient Episodic Memory (GEM)~\cite{lopez-pazGradientEpisodicMemory2017}. GEM stores a subset of data from previous tasks in episodic memory and uses it to constrain the gradient updates during training. However, the server has no training data to store. Thus, we propose to alter the primitive GEM to fit the Federated Learning scenario.

Considering the constraints that the only messages the Server can get are the aggregated models or gradients, different from GEM, we propose to store the gradients instead of data from prior tasks as memory. In FedTracker, we introduce \emph{global memory}. Global memory is defined as the accumulated global gradients from the first iteration to the current iteration. We do not use the global gradient in the current iteration because the gradient of a single iteration tends to be unstable in FL. The global memory can be formally defined as follows.

\begin{definition}[Global memory]
  Global memory is defined as the accumulated global gradients from the first iteration to the current iteration. Global memory $m^t$ of the $t$-th iteration can be calculated using Equation~(\ref{eq:gm}), 
  \begin{equation}
    \label{eq:gm}
    m^t=\sum_{j=1}^{t}G^j=\sum_{j=1}^{t}\sum_{i=1}^K g_i^j,
  \end{equation}
  where $g_i^j$ is the local gradients of the $i$-th Client in $j$-th iteration and $G^j$ is the aggregated global gradients. Because the server cannot get the local gradients of each Client. The server can calculate the global memory by summing up the aggregated gradients $G^j = M_g^j - M_g^{j-1}$ of each iteration.
\end{definition}




The global memory $m^t$ represents the descent direction of the loss of the primitive task. While learning the trigger set task using gradient descent, we should avoid the loss of the primitive task increasing. This can be considered as an inequality constraint measured by computing the angle between the gradients $g$ of the new task and the memory $m^t$. Mathematically, we phrase the optimization problem as:
\begin{equation}
  \min loss(M_g^t, D_T),\enspace s.t.\enspace \langle g, m^t \rangle \geq 0,
\end{equation}
where $g=\frac{\partial loss(M_g^t, D_T)}{\partial M_g^t}$.

If the constraint is satisfied, we can directly apply the gradients $g$ to optimize the loss of the watermark task without increasing the loss of the primitive task. However, if the violation occurs, we should project the gradients $g$ to the closest point $\tilde{g}$ satisfying the constraint. The projection problem can be phased as Equation~(\ref{eq:optim}). The problem can be solved via Quadratic Programming.
\begin{equation}
  \label{eq:optim}
  \begin{aligned}
    \mathop{\arg \min}\limits_{\tilde{g}}\left\lVert g-\tilde{g} \right\rVert_2^2, \enspace s.t.\enspace \langle \tilde{g}, m^t \rangle \geq 0.
  \end{aligned}
\end{equation}

\subsection{Distinguish Different Models of Different Clients}
\label{sec:distinguish}

As discussed before, a key challenge in tracing the model leaker in FL is that the aggregated models received by the Clients are identical. We need to design a mechanism to embed unique, unremovable, and distinguishable identity certificates in the local models. Thus, in FedTracker, we propose the local fingerprint mechanism.

Local fingerprint is used to represent the identity of the Clients. In FedTracker, we utilize the multi-bit \emph{parameter-based} method as local fingerprints. A unique local fingerprint is inserted into the model of each Client. The local fingerprint mechanism includes three phases: local fingerprint generation, local fingerprint insertion, and tracing. 

\vspace{0.3em}
\noindent\textbf{Local Fingerprints Generation.} While generating the fingerprint $F_i$, we propose to maximize the distance between fingerprints to better distinguish them. The bigger the gap between different fingerprints, the more accurate it is for the FL group to disclose the identity of the model leaker among the FL participants. We model the insight as an optimization problem. In FedTracker, we use Hamming distance as a metric since the fingerprint values are discrete in the generation phase. We need to maximize the minimal hamming distance among the fingerprints. The optimization problem can be defined as Equation (\ref{eq:fpoptim}).

\begin{equation}
  \label{eq:fpoptim}
  \{F_i\}_{i=1}^K=\arg\max\min_{1\leq i,j\leq K}{\tt HD}(F_i, F_j),
\end{equation}
where ${\tt HD}(F_i, F_j)$ calculates the Hamming distance between the two fingerprints $F_i$ and $F_j$.

However, the optimization problem is an NP-hard problem, which has no known polynomial-time solution. We therefore utilize the genetic algorithm (GA)~\cite{lambora2019genetic} to solve the optimization problem and generate the fingerprints.

\vspace{0.3em}
\noindent \textbf{Local Fingerprint Insertion.} In FedTracker, given a weight matrix $W_i \in  \mathbb{R}^M$ of the Client $i$'s model, we use the following equation to represent the local fingerprint $F\in \{-1, 1\}^{N}$
\begin{equation}
  \label{eq:lf}
  {\tt sgn}(AW)=F,
\end{equation}
where $A_i$ is the secret key matrix that is randomly generated by the Server. In FedTracker, the secret key matrix $A_i \in \mathbb{R}^{M\times N}$ is generated for each fingerprint $F_i\in \{-1, 1\}^{N}$. ${\tt sgn}(\cdot)$ originates from the sign function.
\begin{equation}
  {\tt sgn}(x)=
  \left\{
  \begin{aligned}
    -1, x<0\\
    1, x\geq 0
  \end{aligned}
  \right..
\end{equation}

In FedTracker, we choose the weight of the BN layers to insert local fingerprints. 
Recall that in BN layers, we calculate the following equation~\cite{ioffe2015batch} to normalize the output $x$ of the previous layer,
\begin{equation}
  o=\gamma\frac{x-\mu}{\sqrt{\sigma^2 + \epsilon }}+ \beta,
\end{equation}
where $\mu$ and $\sigma$ are the mean and variance of input batch $x$, $\epsilon$ is a small constant. $\gamma$ and $\beta$ are learnable parameters. We choose $\gamma$ to embed the local fingerprint. $\gamma$ from different BN layers are concatenated into one vector $W^\gamma$. We define the Hinge-like loss function $L_{f}(W^\gamma_i)$ of fitting the fingerprint for the $i$-th Client as Equation (\ref{eq:lfloss}).
\begin{equation}
  \label{eq:lfloss}
  L_{f}(W_i^\gamma)={\tt HL}(A_i, F_i, W_i^\gamma)=\sum_{j=1}^{|F_i|}\max(\delta - b_{ij}f_{ij}, 0).
\end{equation}

In Equation (\ref{eq:lfloss}), $b_{ij}$ is the $j$-th bit in $B_i=A_iW^\gamma_i$, while $f_{ij}$ is the $j$-th bit in $F_i$. $\delta$ is the hyperparameter controlling the robustness of the fingerprint. We optimize $L_f$ using gradient descent with a small learning rate to reduce the impact on other tasks.

\vspace{0.3em}
\noindent \textbf{Fingerprints Similarity Score.} After inserting the local fingerprints, it is necessary to define a metric to measure the similarity between the extracted fingerprint and the Clients' fingerprints. Existing works mainly utilize Hamming distance to measure the difference between two fingerprints~\cite{liFedIPROwnershipVerification2022,uchida2017embedding}. However, Hamming distance is a metric whose outputs are discrete. There may be ambiguity when finding the most similar fingerprint. Specifically, for an extracted fingerprint $\hat{F}$, there are probably two fingerprints $F_i, F_j$ that ${\tt HD}(\hat{F}, F_i)={\tt HD}(\hat{F}, F_j)$, since the hamming distance rounds the distance to integer. Using Hamming distance is not feasible in our scenario.

In FedTracker, we consider defining a new metric whose outputs are continuous to measure the difference between fingerprints better. Therefore, we propose the Fingerprint Similarity Score:
\begin{definition}[Fingerprint Similarity Score]
  Given the secret key matrix $A_i$, the weight matrix $W$, the fingerprint $F_i$ and $B_i=A_iW$, the Fingerprint Similarity Score (FSS) for the $i$-th Client can be calculated with Equation (\ref{eq:fss}).
  \begin{equation}
    \label{eq:fss}
    {\tt FSS}(A_i, F_i, W) = {\tt FSS}(B_i, F_i)=\sum_{j=1}^{|F_i|}\min(\delta, b_{ij}f_{ij}).
  \end{equation}
\end{definition}

FSS can provide a continuous metric and better robustness for traceability. The further experiments comparing the two metrics are shown in Section \ref{sec:comparing}. The traditional Hamming distance rounds the values and reduces the differentiation. While tracing the leaker, we calculate the FSS for every $A_i$ and $F_i$, and get a similarity vector. The Client with the maximal FSS is suspected of being the leaker.

\section{Implementation of FedTracker}
\label{sec:implementation}

In this section, we provide a comprehensive account of the detailed workflow and implementation of FedTracker. Generally speaking, FedTracker can be defined as a tuple $({\tt Gen, GEmbed, LInsert, Verify, Trace})$.

\subsection{Watermark and Fingerprint Generation}
\label{sec:gen}

In FedTracker, the server should first generate the global watermark for ownership verification and the local fingerprints for traceability. The watermark and fingerprint generation algorithm is abstracted to ${\tt Gen}(\cdot)\rightarrow (D_T, \{F_i\}_{i=1}^K)$. The output of ${\tt Gen}(\cdot)$ has two parts. The former is the global watermark $D_T$, and the latter is $K$ different local fingerprints which are denoted as the set $\{F_i\}_{i=1}^K$. The pseudocode of the watermark and fingerprint generation algorithm is demonstrated in Algorithm \ref{algo:gen}.

The global watermark is a specific trigger set $D_T$, which is composed of the trigger set data $x^T$ and the trigger set labels $y^T$. Any data-free trigger set generation algorithm can be utilized in the implementation.

\begin{figure}[t]
\vspace{-8pt}
\begin{algorithm}[H]
  \caption{Watermark and Fingerprints Generation ${\tt Gen}(\cdot)$}
  \label{algo:gen}
  \hspace*{0.05in}{\bf Input:} The length of fingerprints $N$, the length of embedding layers' weights $M$. \\
  \hspace*{0.05in}{\bf Output:} The watermark trigger set $D_T$, the local fingerprints $\{(F_i, A_i)\}_{i=1}^K$.
  \begin{algorithmic}[1]
  \State $D_T = {\tt WMGen}()$ // \emph{Generate the trigger set.}
  \For{$i=1$ to $K$}
    \State $A_i = {\tt randn}((M, N))$. // \emph{Sample from standard Gaussian distribution.}
    \State $F_i = {\tt choice}(N, [-1, 1])$. // \emph{Initialize the local fingerprint.}
  \EndFor
  \State $\{F_i\}_{i=1}^K = {\tt GASolve}(\{F_i\}_{i=1}^K, {\tt HD})$. // \emph{Solve the optimization problem via GA.}
  \State \Return $D_T, \{(F_i, A_i)\}_{i=1}^K$.
\end{algorithmic}
\end{algorithm}
\vspace{-15pt}
\end{figure}

The local fingerprint for each Client also consists of two parts, a secret key matrix $A_i \in \mathbb{R}^{M\times N}$ and a fingerprint bit string $F_i\in \{-1, 1\}^N$. In our implementation, we randomly generate the secret key matrix $A_i$ from the standard Gaussian distribution $\mathcal{N}(0, 1)$. We generate the fingerprints by solving the optimization problem in Section \ref{sec:distinguish}. 

\subsection{Global Watermark Embedding}
\label{sec:gw}

The global watermark aims to provide the function of ownership verification for FL models. The global watermark embedding algorithm ${\tt GEmbed}(M_g, D_T)\rightarrow \hat{M}_g$ is to embed the global watermark $D_T$ into the model $M_g$, and output the watermarked model $\hat{M_g}$. By watermarking the global model before inserting the Client-specific local fingerprints, all the models in FL are protected by the global watermark. The pseudocode of the global watermark embedding algorithm is shown in Algorithm \ref{algo:gw}.

In each iteration of the global watermark embedding algorithm, the server first updates the global memory using Equation (\ref{eq:gm}). Subsequently, the BN layers of the global model are frozen by the server, implying that parameters associated with these layers remain unaltered during retraining. FedTracker then computes the global gradients and projects them to the direction of the global memory $m^t$ through Equation (\ref{eq:optim}). Additionally, we employ an early stopping technique to preserve utility in our FL model. The retraining process concludes when accuracy on the watermark trigger set surpasses a threshold $\tau_w$.

As demonstrated in Algorithm \ref{algo:gw}, we freeze the BN layers in the model while learning the trigger set task on account of the following two reasons. First, as analyzed in Section~\ref{sec:embed}, the trigger set generated by the data-free trigger set construction method~\cite{tekgul2021waffle, adi2018turning} is out of the distribution of natural data. Directly learning the trigger set can significantly affect the stability of BN layers and decrease the utility. Thus freezing the BN layers can help retain the utility of the model. Second, Freezing the BN layers can also avoid negative influence on local fingerprints, since the local fingerprints are embedded into the BN layers.

\begin{figure}[t]
\vspace{-8pt}
\begin{algorithm}[H]
  \caption{Global Watermark Embedding ${\tt GEmbed}(\cdot)$}
  \label{algo:gw}
  \hspace*{0.05in}{\bf Input:} The watermark trigger set $D_T$, the aggregated global model $M^t_g$ in $i-$th iteration, the global watermark retraining learning rate $\lambda_w$ \\
  \hspace*{0.05in}{\bf Output:} The watermarked global model $\hat{M}^t_g$.
  \begin{algorithmic}[1]
  \State $m^t = \frac{1}{t}m^{t-1} + \frac{t-1}{t}(M^t_g - M^{t-1}_g)$ // \emph{Update global memory.}
  \State $i=0$
  \State $\hat{M}^t_g=M^t_g$
  \State ${\tt FreezeBN}(\hat{M}^t_g)$
  \While{$i<{\tt max\_iter}$ and ${\tt acc}(\hat{M}^t_g, D_T) \leq \tau_w$}
    \State $i=i+1$
    \State $g=\frac{\partial loss(\hat{M}_g^t, D_T)}{\partial \hat{M}_g^t}$
    \State $\tilde{g}={\tt project}(g, m^t)$
    \State $\hat{M}^t_g=\hat{M}^t_g - \lambda_w\tilde{g}$
  \EndWhile
  \State \Return $\hat{M}^t_g$.
\end{algorithmic}
\end{algorithm}
\vspace{-15pt}
\end{figure}

\subsection{Local Fingerprints Insertion}
\label{sec:lf}

The local fingerprints aim to identify the models of Clients and trace the model leaker via the fingerprint inside the suspicious model. The local fingerprint insertion algorithm ${\tt LInsert}(\hat{M}_g, F_i)\rightarrow \hat{M_i}$ takes the watermarked model $\hat{M_g}$ and the Client's fingerprint $F_i$ as input, and output the model $\hat{M_i}$. $\hat{M_i}$ contains both the global watermark and the local fingerprint. The pseudocode of local fingerprint insertion can be found in Algorithm \ref{algo:lf}. 

In each iteration of the local fingerprint insertion algorithm, the server first makes a copy of the watermarked global model and gets the weights $W_i$ of the layers. Subsequently, employing the gradient descent algorithm, the server optimizes the fingerprint loss as defined in Equation~(\ref{eq:lfloss}). Zheng et al.~\cite{zheng2022DNNFingerprintNonRepudiable} reported that the entropy of the weights from the front layers is higher than those from the deeper layers. Thus, fingerprints inserted into the deeper layers are less likely to be erased by aggregation. We choose the deeper BN layers to insert the fingerprints.

In FedTracker, we should avoid the situation that the model simultaneously overfits on two fingerprints, $i.e.$, for one suspicious model $\hat{M}_{adv}$, there are two fingerprints $F_\mu, F_\nu$ having the same maximal FSS. This can lead to ambiguity and make the tracing fail. Li et al.~\cite{liFedIPROwnershipVerification2022} has proven that there exists a $W$ containing all the fingerprints if $KN \leq M$. $N$ and $M$ are the dimensions of $F$ and $W$. Therefore, while inserting the local fingerprints, we should follow $KN > M$. FedTracker also utilizes an early-stopping technique to prevent the situation. When the FSS is higher than a threshold $\tau_f$, FedTracker will stop inserting the fingerprint. It can also reduce the influence of inserting fingerprints on the utility of the model.

\subsection{Verification and Tracing}
\label{sec:verify}

Once the FL group suspects that the FL model is deployed by an unauthorized party, the FL group should conduct the ownership verification and traceability mechanisms to protect their copyright and catch the culprit. The overview of ownership verification and traceability mechanisms are demonstrated in Fig.~\ref{fig:problem}.

\vspace{0.3em}
\noindent\textbf{Ownership Verification:} The watermark verification algorithm ${\tt Verify}(\hat{M}_{adv}, D_T)$ confirms the existence of the watermark in the suspicious model. If the watermark can be extracted from the suspicious model, the FL group can successfully verify their ownership. The accuracy of the specific trigger set indicates the existence of the watermark. 
\begin{equation}
  \label{eq:verify}
  {\tt Verify}(\hat{M}_{adv}, D_T)=\left\{
  \begin{aligned}
    &{\tt True}, {\tt acc}(\hat{M}_{adv}, D_T) \geq \epsilon_v \\
    &{\tt False}, {\tt if\enspace otherwise}
  \end{aligned}
\right.,
\end{equation}
where $\hat{M}_{adv}$ is the model from the adversary. As shown in the middle part of Fig. \ref{fig:problem}, the FL group uses the watermark token, $i.e.$, the trigger set to query the API of the suspicious model. Then the FL group checks the output and calculates the accuracy of the trigger set. As shown in Equation (\ref{eq:verify}), if the accuracy is greater than the preset threshold $\epsilon_v$, the ownership can be verified.

\begin{figure}[t]
\vspace{-10pt}
\begin{algorithm}[H]
  \caption{Local Fingerprint Insertion ${\tt LInsert}(\cdot)$}
  \label{algo:lf}
  \hspace*{0.05in}{\bf Input:} The local fingerprint $F_i$ of $i-$th Client, the secret key matrix $A_i$, the local fingerprint insertion learning rate $\lambda_f$, the watermarked global model $\hat{M}^t_g$\\
  \hspace*{0.05in}{\bf Output:} The local model $\hat{M}^t_i$ with the fingerprint.
  \begin{algorithmic}[1]
  \State $\hat{M}_i^t={\tt Copy}(\hat{M}^t_g)$
  \State $W_i={\tt GetLayerWeight}(\hat{M}_i^t)$
  \State $j=0$
  \While{$j<{\tt max\_iter}$ and ${\tt FSS}(A_i, F_i, W_i) \leq \tau_f$}
    \State $j=j+1$
    \State $g=\frac{\partial {\tt HL}(A_i, F_i, W_i)}{\partial \hat{M}_g^t}$
    \State $W_i=W_i-\lambda_f g$
  \EndWhile
  \State \Return $\hat{M}^t_i$.
\end{algorithmic}
\end{algorithm}
\vspace{-15pt}
\end{figure}

\vspace{0.3em}
\noindent\textbf{Traceability:} After successfully verifying the ownership, the leaker tracing algorithm ${\tt Trace}(\hat{M}_{adv}, \{F_i\}_{i=1}^K)$ extracts the fingerprint $F_{adv}$ from the suspicious model $\hat{M}_{adv}$, and then check the most similar fingerprint among the Clients' fingerprints set $\{F_i\}_{i=1}^K$. As shown in the right part of Fig. \ref{fig:problem}, the FL group compares the fingerprint with each Client's fingerprint and finds the most similar fingerprint. The traceability mechanism outputs the index of the malicious Client by Equation~(\ref{eq:fsstrace}).
\begin{equation}
  \label{eq:fsstrace}
  {\tt Trace}(\hat{M}_{adv}, \{F_i\}_{i=1}^{K})=\arg\max_j {\tt FSS}(A_jW_{adv}, F_j),
\end{equation}
where $W_{adv}$ is the weight of the fingerprint embedding layer in the model $\hat{M}_{adv}$. By utilizing the designed leaker tracing algorithm, any suspicious model can be traced back to its owner inside the FL group. Since each Client's model is assigned with a distinct fingerprint, FedTracker can find the malicious leaker despite the number of malicious Clients.

\section{Experiments}
\label{sec:experiment}

\subsection{Experiment Settings}

\noindent\textbf{Datasets and Models.} In our experiments, we use four different Convolution Neural Networks (CNN) to learn three different datasets including MNIST~\cite{y2010MNIST}, CIFAR10, and CIFAR100~\cite{krizhevsky2009learning}. MNIST is a widely used benchmark dataset with 60,000 gray images. CIFAR10 and CIFAR100 consist of 50,000 colored images for training and 10,000 for testing. To keep consistency, we resize the images in all the datasets to $32\times 32$ in our implementation. 

For MNIST, following prior works~\cite{mcmahan2017communication}, we utilize a four-layer CNN (denoted as CNN-4) which is of the LeNet~\cite{lecun1998gradient} style. For CIFAR10, two popular models, AlexNet~\cite{krizhevsky2017imagenet} and VGG-16~\cite{simonyan2014very}, are chosen. ResNet-18~\cite{he2016deep} is trained on CIFAR100, a dataset with 100 classes. The first three models are slightly modified by adding a BN layer after each convolution layer since it is proven that adding BN layers can achieve better performance.

\vspace{0.3em}
\noindent\textbf{FL Settings.} To simulate the FL scenario, we set the default number of Clients to $50$. In each iteration, we choose $40\%$ Clients to train their local models for $2$ local epochs. The learning rate of Clients is set to $0.01$. For data distribution of different FL Clients, we evaluate FedTracker under both independent identical distribution (i.i.d) setting and none independent identical distribution (non-i.i.d) setting in Section~\ref{sec:iid} and Section \ref{sec:noniid} respectively.

\vspace{0.3em}
\noindent\textbf{Watermark and Fingerprint.} For the global watermark, we implement WafflePattern~\cite{tekgul2021waffle} to generate the trigger set. We first choose a pattern for each class and add Gaussian noise to the patterns to generate the trigger set samples. Each class has $10$ samples in the trigger set. The learning rate is set to $0.005$ while embedding global watermarks. For local fingerprints, we set the default length of fingerprints to be $128$, which is sufficient for tracing $50$ Clients. The learning rate is $0.05$ when inserting the Client's fingerprint. 

\begin{table}[t]
  \caption{Watermark accuracy (\%) under different numbers of Clients.}
  \label{table:wmc}
  \centering
  \setstretch{1.1}
  \small
  \begin{tabular}{c|cccc}
  \hline
  \hline
   & \multicolumn{4}{c}{Number of Clients} \\
  Model & 10 & 20 & 30 & 50\\
  \hline
  CNN-4 on MNIST & 84.70 & 87.80 & 80.70 & 86.90\\
  AlexNet on CIFAR10 & 98.90 & 96.45 & 86.83 & 82.68\\
  VGG-16 on CIFAR10 & 99.30 & 99.90 & 99.87 & 99.12\\
  ResNet-18 on CIFAR100 & 99.60 & 98.45 & 90.70 & 83.06\\
  \hline
  \hline
  \end{tabular}
  \vspace{-10pt}
\end{table}

\begin{table}[t]
  \caption{Watermark accuracy (\%) under different sizes of watermark trigger set.}
  \label{table:wms}
  \centering
  \setstretch{1.1}
  \small
  \begin{tabular}{c|ccccc}
  \hline
  \hline
  & & \multicolumn{4}{c}{Size of watermark trigger set} \\
  Model & Metric & 100 & 200 & 300 & 500\\
  \hline
  \multirow{2}{*}{AlexNet} & WM acc & 82.68 & 77.26 & 81.86 & 79.89\\
  & Test acc & 84.32 & 84.06 & 84.10 & 84.07\\
  \hline
  \multirow{2}{*}{VGG-16} & WM acc & 99.12 & 99.23 & 97.52 & 88.50\\
  & Test acc & 88.63 & 88.61 & 88.60 & 88.56 \\
  \hline
  \hline
  \end{tabular}
  \vspace{-15pt}
\end{table}

\vspace{0.3em}
\noindent\textbf{Metrics.} To evaluate the utility of the FL models, we calculate the accuracy on a 10,000-sample testing set as the metric. For ownership verification, we use the accuracy of the trigger set, named watermark accuracy (WM acc for short), as the metric. For traceability, we define another metric named \emph{Traceability Rate}. Traceability Rate (TR) is defined as the rate at which the Client with the largest FSS happens to be the holder of the model. In all our experiments, FedTracker achieves a $100\%$ traceability rate, which indicates that FedTracker can always find the actual owner of the suspicious model.

\begin{definition}[Traceability Rate]
    Traceability Rate (TR) is defined as the rate at which the Clients with the largest FSS are actually the original owner of the model. TR can be computed using Equation (\ref{eq:tr}).
    \begin{equation}
      \label{eq:tr}
      {\tt TR} = (\sum_{i=1}^K [\mathop{\arg \max}\limits_{j} {\tt FSS}(A_j, F_j, W_i)] = i) / K.
    \end{equation}
\end{definition}



\subsection{Experiments under i.i.d Settings}
\label{sec:iid}

Based on the design objectives listed in Section \ref{sec:defender}, FedTracker needs to achieve four basic properties: ownership verification, traceability, fidelity, and robustness. In this section, we evaluate the performance of FedTracker under i.i.d settings. We assume that the distribution of the Clients' data follows i.i.d, which means the training dataset is uniformly allocated to each Client. The i.i.d setting is the most common setting in FL.

\subsubsection{Effectiveness on Ownership Verification}
\label{sec:eff1}

In this subsection, we implement FedTracker in different settings to evaluate the effectiveness on ownership verification, $i.e.$, whether the global watermark is successfully embedded into every model in FL.

\begin{figure}[t]
  \centering
  \includegraphics[width=0.95\linewidth]{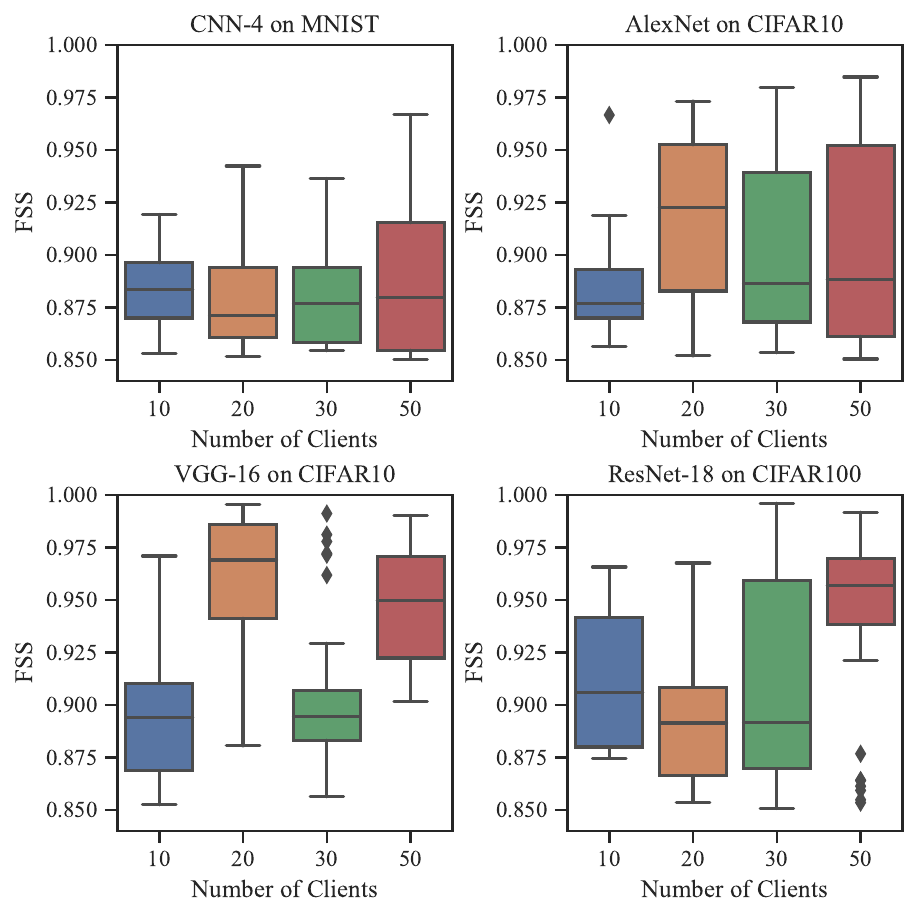}
  \caption{FSS evaluation under different numbers of Clients.}
  \label{fig:ffsc}
  \vspace{-10pt}
\end{figure}

\begin{figure}[t]
  \centering
  \includegraphics[width=0.95\linewidth]{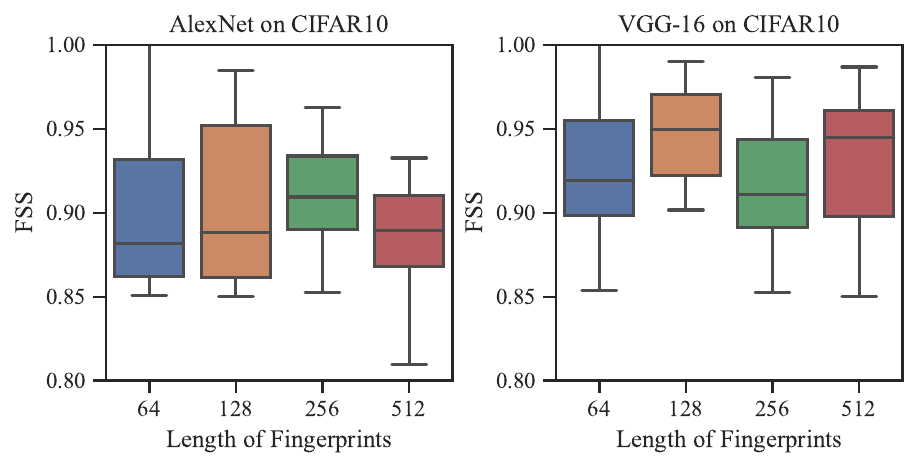}
  \caption{FSS evaluation under different lengths of local fingerprints.}
  \label{fig:fssl}
  \vspace{-15pt}
\end{figure}

\vspace{0.3em}
\noindent\textbf{Ownership Verification under Different Numbers of the Clients.} 
We implement FedTracker with four scenarios with different numbers of Clients in FL, $10, 20, 30$, and $50$. For ownership verification, the results of the average watermark accuracy are shown in Table \ref{table:wmc}. As the number of Clients grows, the watermark accuracy decreases to some extent. But in all the cases, the watermark accuracy is still greater than $80\%$, which indicates that the watermark is successfully embedded into the model.

\vspace{0.3em}
\noindent\textbf{Ownership Verification under Different Sizes of Watermark Trigger Set.} The size of the watermark trigger set is defined as the number of samples in the trigger set. In this experiment, we evaluate the ownership verification effectiveness under different sizes of the watermark trigger set. We generate $100, 200, 300$, and $500$ samples to construct the trigger set. The results in Table \ref{table:wms} demonstrate that with the size of the trigger set increasing, both the watermark accuracy and the testing accuracy drops slightly. However, the watermark accuracy is still high enough for ownership verification, which proves that the watermark is retained in the FL models. The utility degradation is also negligible. It demonstrates that FedTracker is effective despite the size of the trigger set.

\subsubsection{Effectiveness on Traceability}

We further evaluate the effectiveness of FedTracker on traceability under various settings, $i.e.$, whether the inserted local fingerprints can identify the specific Client. We implement FedTracker with different numbers of Clients and different lengths of local fingerprints.

\vspace{0.3em}
\noindent\textbf{Traceability under Different Numbers of Clients.} Under the same settings in Section \ref{sec:eff1}, for traceability, we achieve $100\%$ TR in all experiments. The FSS results are shown in Fig. \ref{fig:ffsc}. In the experiments, the lowest FSS is higher than $0.80$, which indicates a successful insertion of local fingerprints. This experiment also indicates that FedTracker can accurately trace the source Client even with a large number of (malicious) Clients.

\begin{figure*}[t]
  \centering
  \includegraphics[width=0.95\linewidth]{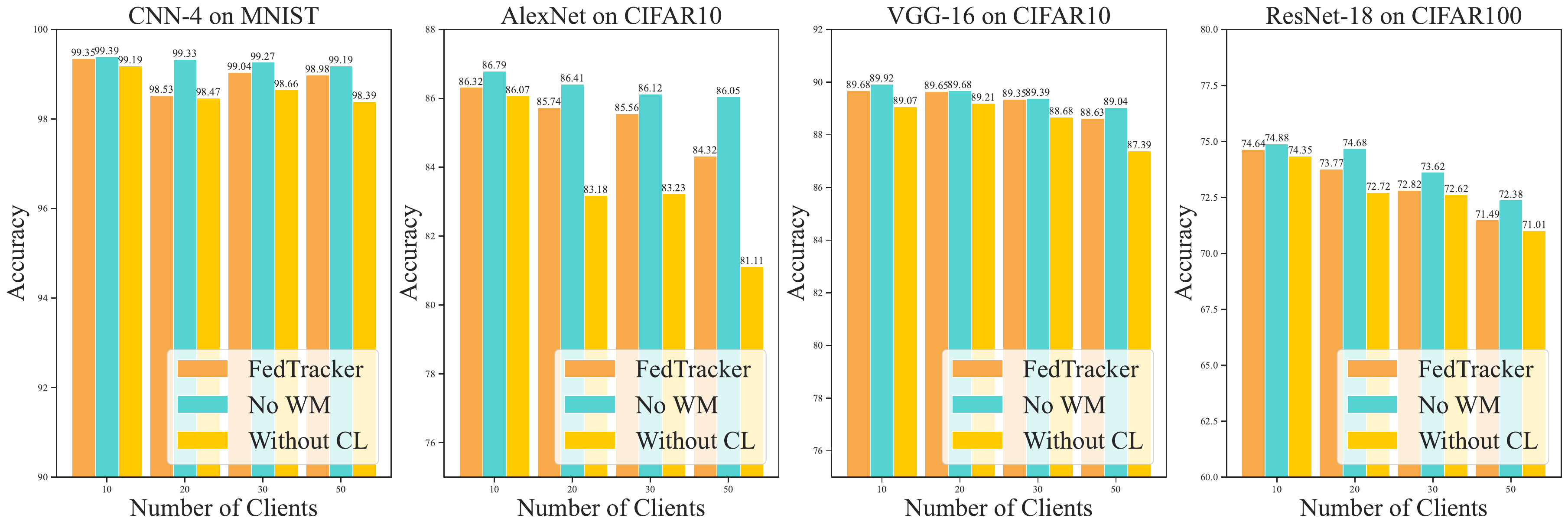}
  \caption{Accuracy on the primitive task of FedTracker comparing with the model without watermark (No WM) and without CL.}
  \label{fig:fide}
  \vspace{-10pt}
\end{figure*}

\vspace{0.3em}
\noindent\textbf{Traceability under Different Lengths of Local Fingerprints.} In this experiment, we test the traceability effectiveness of FedTracker by setting the fingerprint length to $64, 128, 256,$ and $512$ bits. The number of Clients is fixed at $50$ Clients, which is the most complicated case in our simulation. We still achieve $100\%$ TR, and the FSS evaluations are shown in Fig.~\ref{fig:fssl}. Most of the FSS metrics are around $0.90$. Experimental results demonstrate that the length of the local fingerprints does not significantly affect the traceability effectiveness, since the minimal FSS is greater than $0.80$. 

\subsubsection{Fidelity}

Fidelity signifies that embedding watermarks and fingerprints does not have a significant impact on the utility of the FL models. In the fidelity evaluations, we first compare the utility of the protected model with the model trained without any watermark embedding and fingerprint insertion. The utility of the model can be evaluated by the accuracy of the primitive task, $i.e.$, the accuracy of the validation set. As shown in Fig. \ref{fig:fide}, FedTracker can achieve a similar utility compared with the model without a watermark. The accuracy degradations in all the experiments are within $1\%$, which signifies a good fidelity of FedTracker.

We also conduct an ablation study of the CL-based watermark embedding method in FedTracker. The results are also demonstrated in Fig. \ref{fig:fide}. It shows that using our CL-based embedding algorithm can improve the utility of the model with at most $3.21\%$ testing accuracy improvement. In all the experiments, the utility using CL is better than without CL, which indicates the effectiveness of our CL-based algorithm.

\subsubsection{Robustness}
\label{sec:robustness}

Robustness indicates whether the watermarks and the fingerprints can be removed by the adversaries. In the robustness experiments, we evaluate the robustness of the global watermark and local fingerprints against several watermark removal attacks which are introduced in Section~\ref{sec:threat}. Given a specific attack method ${\tt att}(\cdot)$, the watermark or fingerprints are considered robust if one of the following two cases is true after the attack:
\begin{case}
  \label{case:robustness1}
  This case refers to that the watermark and fingerprints are not significantly changed by the attack. The ownership verification and traceability mechanisms can still work properly. It can be defined as Equation (\ref{eq:robustness1}).
  \begin{equation}
    \label{eq:robustness1}
    \begin{aligned}
      &[{\tt Verify}({\tt att}(\hat{M}_{adv}),D_T)={\tt True} \\
      \wedge\enspace & {\tt Trace}({\tt att}(\hat{M}_{adv}),\{F_i\}_{i=1}^K)= adv]
    \end{aligned}
  \end{equation}
\end{case}
\begin{case}
  \label{case:robustness2}
  This case signifies that the attack has succeeded in removing the watermark and fingerprints. However, the functionality of the attacked model drops significantly compared to the primitive model and is no longer useful for its task. This case also renders an unsuccessful attack. Equation (\ref{eq:robustness2}) defines the case formally.
  \begin{equation}
    \label{eq:robustness2}
    \begin{aligned}
      &[{\tt Verify}({\tt att}(\hat{M}_{adv}),D_T)={\tt False} \\
      \vee\enspace & {\tt Trace}({\tt att}(\hat{M}_{adv}),\{F_i\}_{i=1}^K)\neq adv]\\ 
      \wedge\enspace& {\tt Acc}(\hat{M}_{adv}) - {\tt Acc}({\tt att}(\hat{M}_{adv})) > \tau
    \end{aligned}
  \end{equation}
\end{case}

\begin{figure}[t]
  \centering
  \includegraphics[width=0.95\linewidth]{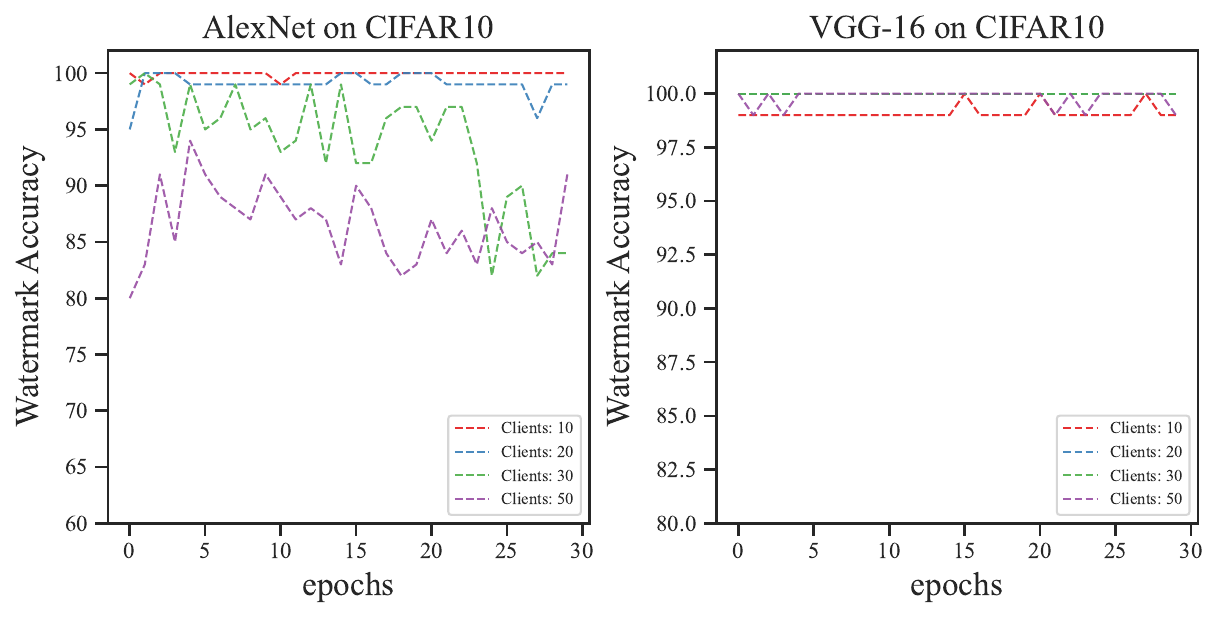}
  \caption{Watermark accuracy (WM) against fine-tuning attack. The number behind refers to the number of Clients in FL.}
  \label{fig:finetune-wm}
  \vspace{-10pt}
\end{figure}

\begin{figure}[t]
  \centering
  \includegraphics[width=0.95\linewidth]{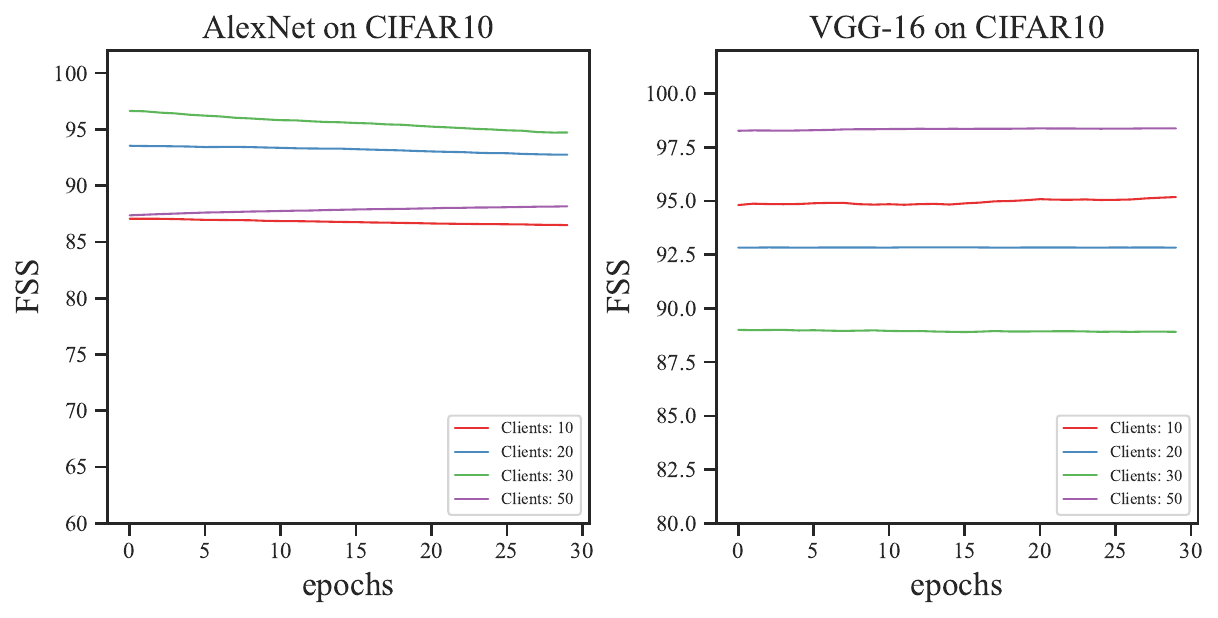}
  \caption{FSS against fine-tuning attack. The number behind refers to the number of Clients in FL.}
  \label{fig:finetune-fss}
  \vspace{-10pt}
\end{figure}

\vspace{0.3em}
\noindent\textbf{Robustness against Fine-tuning Attack.} Fine-tuning attack refers to training the model with the private dataset of the malicious Client for several epochs. We randomly choose a Client to conduct a $30$-epoch fine-tuning. In Fig. \ref{fig:finetune-fss}, the FSS metric rarely changes in the $30$ epochs. While the watermark accuracy fluctuates as shown in Fig. \ref{fig:finetune-wm} but remains a high value in general, which indicates the robustness of FedTracker against fine-tuning.

\begin{table}[t]
  \caption{FSS, watermark accuracy (WM acc), and classification accuracy (Test acc) against quantization attacks.}
  \label{table:quantization}
  \centering
  \setstretch{1.1}
  \footnotesize
  \begin{tabular}{ccccc}
    \hline
    \hline
    & & \multicolumn{3}{c}{Data type} \\
    Model & Metric(\%) & float32 & float16 & int8 \\
    \hline
    \multirow{3}{*}{AlexNet on CIFAR10} & Test acc & 80.06 & 80.06 & 80.04\\
    & WM acc & 84.00 & 84.00 & 84.00\\
    & FSS & 87.32 & 87.32 & 87.32\\
    \hline
    \multirow{3}{*}{VGG-16 on CIFAR10} & ACC & 88.03 & 88.05 & 88.03\\
    & WM & 98.00 & 98.00 & 97.00\\
    & FSS & 98.28 & 98.27 & 98.22\\
    \hline
    \hline
  \end{tabular}
  \vspace{-5pt}
\end{table}

\begin{figure}[t]
  \centering
  \includegraphics[width=0.95\linewidth]{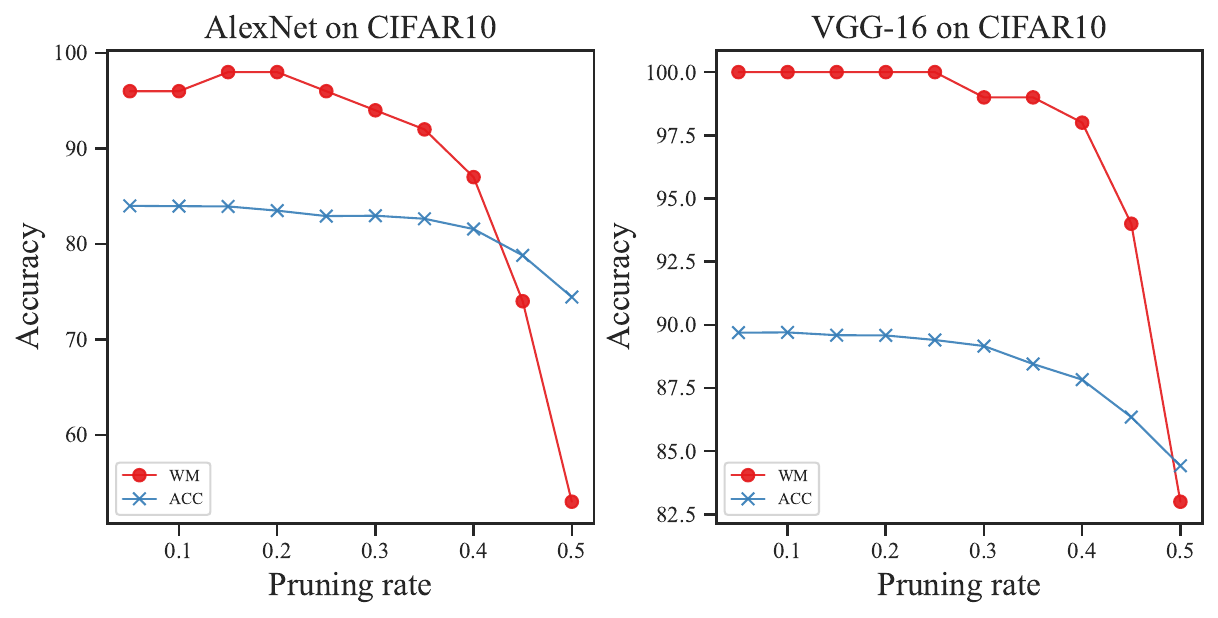}
  \caption{Watermark accuracy (WM) and classification accuracy (ACC) against pruning attack which does not prune the BN layers.}
  \label{fig:pruning}
  \vspace{-10pt}
\end{figure}

\begin{figure}[t!]
  \centering
  \includegraphics[width=0.95\linewidth]{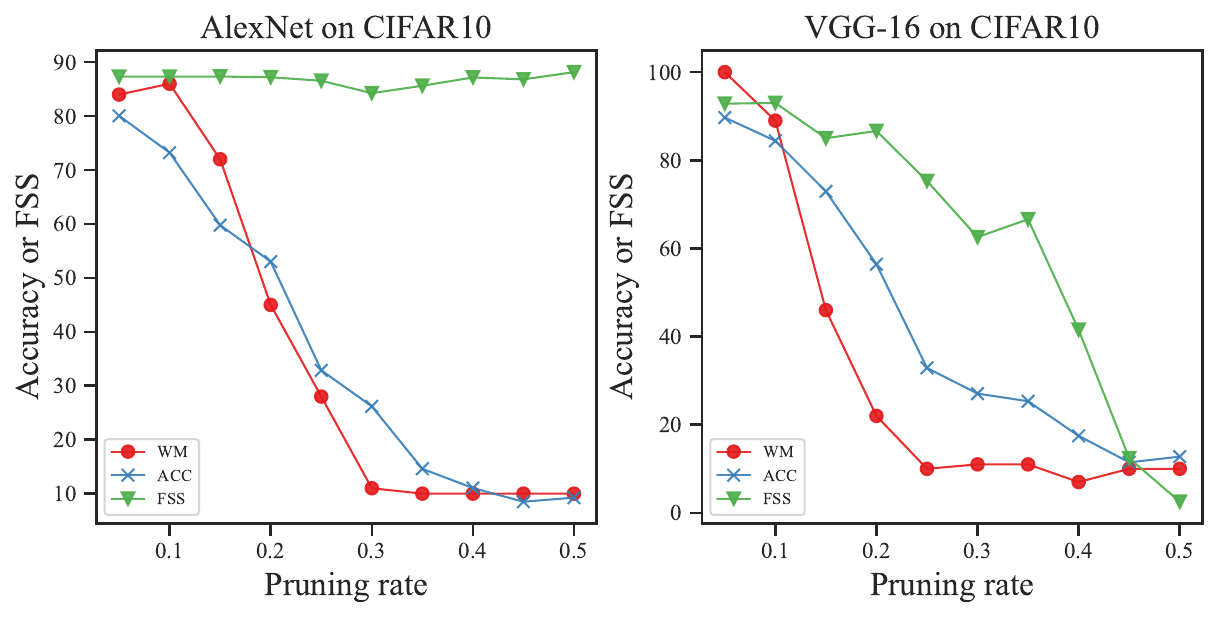}
  \caption{FSS, watermark accuracy (WM) and classification accuracy (ACC) against pruning attack which prunes the BN layers.}
  \label{fig:pruningbn}
  \vspace{-10pt}
\end{figure}

\vspace{0.3em}
\noindent\textbf{Robustness against Quantization Attack.} Quantization attack is the most effective compression technique in practice. The quantization attack refers to using fewer bits to represent the parameters. In the quantization attack experiments, we test three different data types, float32, float16, and int8. The results are shown in Table \ref{table:quantization}.
  
The results demonstrate that although quantization reduces the bits of model parameters, the watermark accuracy and FSS of FedTraker are rarely affected. The results indicate the robustness of FedTraker against the quantization attack.

\vspace{0.3em}
\noindent\textbf{Robustness against Pruning Attack.} We evaluate the robustness of our watermark and fingerprints against pruning attack, especially parameter pruning~\cite{han2015learning}. In parameter pruning, which is also conducted in ~\cite{uchida2017embedding}, we prune units in the model by zeroing out the ones with the lowest $l_1$-norm. We divide the pruning attack into two cases: (1) pruning all the layers of the model except the BN layers, and (2) pruning the BN layers. 

In the first case, pruning other layers does not affect the fingerprints in BN layers. Thus, the watermark is the only one that might be influenced. The watermark accuracies and classification accuracies after pruning are shown in Fig. \ref{fig:pruning}. As the pruning rate increases, the watermark accuracy and classification accuracy decrease simultaneously. In most cases, the watermark accuracy is above $80\%$. The only exception is pruning $50\%$ neurons of AlexNet. However, the classification accuracy of the pruned model has nearly $10\%$ decrease, which means the model has lost its value. Thus, we do not think it is a successful attack.

In the second case, pruning BN layers can affect both watermarks and fingerprints. From Fig. \ref{fig:pruningbn}, it can be observed that the classification accuracy drops significantly when pruning the BN layers. When the model still retains its utility, the watermark accuracy and FSS are still high enough to provide ownership verification and traceability. In Conclusion, in the two cases, the pruning attack cannot successfully remove the watermark and fingerprints in the FL model, which demonstrates the robustness of FedTracker against the pruning attack.

\begin{figure}[t]
  \centering
  \includegraphics[width=0.95\linewidth]{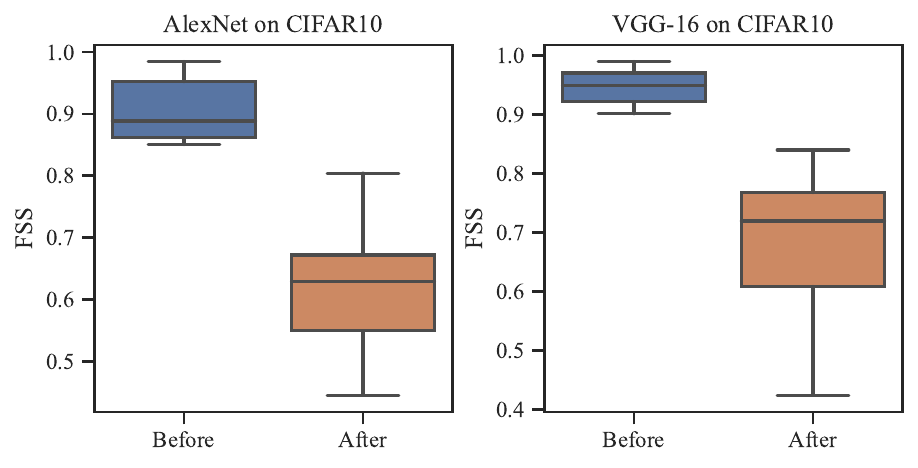}
  \caption{FSS before and after the overwriting attack.}
  \label{fig:overwriting}
  \vspace{-15pt}
\end{figure}

\vspace{0.3em}
\noindent\textbf{Robustness against Overwriting Attack.} The malicious Client might randomly generate a fingerprint $F_m$ and the secret key matrix $A_m$, and try to overwrite the fingerprint in the model. From Fig. \ref{fig:overwriting}, the FSS metrics drop after the overwriting attack. However, the TR of the $50$ models is still $100\%$, which means FedTracker can still trace the leaker successfully. The result demonstrates that although the overwriting attack has an impact on the fingerprints, it has little effect on the relative magnitudes of FSS. FedTracker remains effective against the overwriting attack.

\begin{table*}[t]
   \centering
   \setstretch{1.1}
   \small
   \caption{Testing accuracy, watermark accuracy and FSS (\%) before and after Neural Cleanse.}
   \label{table:nc}
   \begin{tabular}{cc|cc|cc|cc|cc}
   \hline
   \hline
    & & \multicolumn{8}{c}{Number of Clients} \\
    & & \multicolumn{2}{c}{10} & \multicolumn{2}{c}{20} & \multicolumn{2}{c}{30} & \multicolumn{2}{c}{50}\\
    Model & Metric & Before & After & Before & After & Before & After & Before & After \\
   \hline
   \multirow[c]{2}{*}{AlexNet on CIFAR10} & Test acc & 85.89 & 74.39 & 83.99 & 79.73 & 82.74 & 72.79 & 80.06 & 74.58\\
   & WM acc & 100.00 & 72.00 & 96.00 & 77.00 & 98.00 & 80.00 & 84.00 & 69.00\\
   \hline
   \multirow[c]{2}{*}{VGG-16 on CIFAR10} & Test acc & 89.59 & 78.88 & 89.69 & 79.16 & 89.18 & 79.28 & 88.03 & 80.09\\
   & WM acc & 99.00 & 79.00 & 100.00 & 70.00 & 100.00 & 74.00 & 98.00 & 72.00\\
   \hline
   \hline
   \end{tabular}
   \vspace{-10pt}
 \end{table*}

 \begin{table*}[t]
   \centering
   \setstretch{1.1}
   \small
   \caption{Testing accuracy, watermark accuracy before and after FeatureRE.}
   \label{table:featureRE}
   \begin{tabular}{cc|cc|cc|cc|cc}
   \hline
   \hline
   & & \multicolumn{8}{c}{Number of Clients} \\
   & & \multicolumn{2}{c}{10} & \multicolumn{2}{c}{20} & \multicolumn{2}{c}{30} & \multicolumn{2}{c}{50}\\
   Model & Metric & Before & After & Before & After & Before & After & Before & After \\
   \hline
   \multirow[c]{2}{*}{AlexNet on CIFAR10} & Test acc & 85.89 & 70.01 & 83.99 & 71.66 & 82.74 & 77.12 & 80.06 & 73.32\\
   & WM acc & 100.00 & 87.00 & 96.00 & 77.00 & 98.00 & 76.00 & 84.00 & 73.00\\
   \hline
   \multirow[c]{2}{*}{VGG-16 on CIFAR10} & Test acc & 89.59 & 83.41 & 89.69 & 70.90 & 89.18 & 75.30 & 88.03 & 72.71\\
   & WM acc & 99.00 & 87.00 & 100.00 & 84.00 & 100.00 & 88.00 & 98.00 & 77.00\\
   \hline
   \hline
   \end{tabular}
   \vspace{-10pt}
 \end{table*}

\vspace{0.3em}
\noindent\textbf{Robustness against Backdoor Mitigation Methods.} In the backdoor mitigation attack, we implement Neural cleanse~\cite{wang2019neural} and FeatureRE~\cite{wang2022rethinking} trying to remove the watermark inside the FL model. 

Neural cleanse and FeatureRE are two techniques for detecting and removing backdoors in deep neural networks. They both consist of two main steps: trigger reverse engineering and trigger unlearning. Trigger reverse engineering tries to reconstruct the possible triggers that activate the backdoor in a given model. Trigger unlearning then involves retraining the model on the reversed trigger data with correct labels. In this way, the model can forget the association between the trigger and the target class, and restore its normal behavior.

The experimental results of AlexNet and VGG16 trained on CIFAR10 are demonstrated in Table \ref{table:nc} and Table \ref{table:featureRE}. From Table \ref{table:nc}, the neural cleanse method has an impact on the watermark of the model and decreases the watermark accuracy to about $70\%$. However, the testing accuracy also drops about $10\%$, which signifies a significant negative effect on the model utility and value. A similar phenomenon is also found in the FeatureRE method. As shown in Table \ref{table:featureRE}, at the expense of more than $6\%$ testing accuracy degradation, the watermark accuracy is still high enough for verification in most cases.

Results from neural cleanse and FeatureRE indicate the robustness of the watermark against backdoor mitigation methods. The ineffectiveness of the two methods may be ascribed to the principles of backdoor reverse engineering. Existing backdoor mitigation methods use samples from natural datasets to reverse and construct a trigger set, which has a relatively small influence on the trigger set based on noise. A Backdoor based on natural data is more likely to be triggered in the real world and thus more harmful than a backdoor based on noise. It is reasonable for the backdoor mitigation method to mainly focus on removing the former kind of backdoor.

\subsection{Experiments under Non-i.i.d Settings}
\label{sec:noniid}

The non-identically independent distribution (non-i.i.d) of Clients' data is a prevalent issue in FL. Specifically, the dataset held by the Clients in FL exhibits substantial heterogeneity in their distributions, where the labels and the number of samples in the training set of the Clients are significantly different.

Non-i.i.d data has a detrimental impact on the effectiveness of FL as it leads to diverse decision boundaries being learned by each Client's model. Consequently, the global gradient undergoes significant changes during each aggregation iteration. This introduces a potential risk in non-i.i.d settings where the watermark and fingerprints could be erased after aggregation.

Therefore, in this section, we measure the performance of FedTracker under the non-i.i.d setting. Consistent with previous studies~\cite{yurochkin2019bayesian}, we employ the Dirichlet distribution to synthesize non-i.i.d data for Clients in FL. The concentration parameter $\xi \in [0, 1]$ of the Dirichlet distribution governs the extent of non-i.i.dness. A smaller concentration parameter leads to increased divergence among data distributions across different Clients.

\begin{table}[t]
  \caption{Watermark accuracy (\%) under non-i.i.d settings.}
  \label{table:wmn}
  \centering
  \setstretch{1.05}
  \small
  \begin{tabular}{c|ccccc}
  \hline
  \hline
   & \multicolumn{4}{c}{Data distribution} \\
  Model & Metric & i.i.d & $\xi=0.9$ & $\xi=0.7$ & $\xi=0.5$\\
  \hline
  \multirow{2}{*}{AlexNet} & WM acc & 82.68 & 81.90 & 80.24 & 79.50 \\
  & Test acc & 84.32 & 83.52 & 83.24 & 82.95\\
  \hline
  \multirow{2}{*}{VGG-16} & WM acc & 99.12 & 99.14 & 98.40 & 98.88\\
  & Test acc & 88.63 & 88.09 & 87.52 & 87.28 \\
  \hline
  \hline
  \end{tabular}
  \vspace{-5pt}
\end{table}

\begin{figure}[t]
  \centering
  \includegraphics[width=0.95\linewidth]{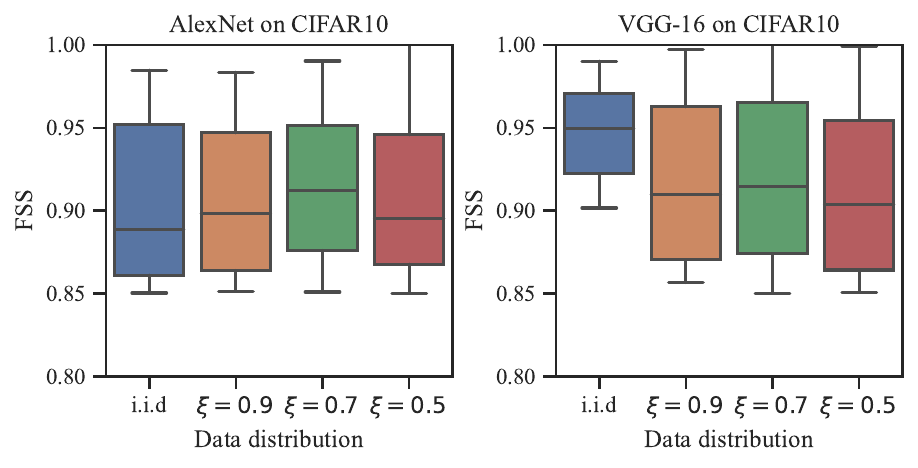}
  \caption{FSS evaluation under non-i.i.d settings.}
  \label{fig:fssn}
  \vspace{-15pt}
\end{figure}

\subsubsection{Effectiveness under Non-i.i.d Settings}
\label{sec:noniid-eff}

We conduct the experiments with $\xi =0.5, 0.7, 0.9$ to evaluate the effectiveness of FedTracker under non-i.i.d settings. The results are demonstrated in Fig. \ref{fig:fssn} and Table \ref{table:wmn}.  It is observed that as the variability among Clients' data increases, both watermark accuracy and testing accuracy decrease simultaneously; however, the watermark accuracy remains above $79\%$. Moreover, the FSS metrics are higher than $0.85$ in all the experiments as depicted in Fig.~\ref{fig:fssn}. The traceability mechanism exhibits minimal impact under non-i.i.d settings as evidenced by negligible changes in FSS metrics.

\begin{table}[t]
    \centering
    \setstretch{1.1}
    \small
    \caption{Testing accuracy of the watermarked model and the models without watermark (No WM) under non-i.i.d settings.}
    \label{table:noniid-fidelity}
    \begin{tabular}{c|cccc}
    \hline
    \hline
    Method & i.i.d & $\xi=0.9$ & $\xi=0.7$ & $\xi=0.5$\\
    \hline
    FedTracker & 88.63 & 87.44 & 87.32 & 87.28\\
    No WM & 89.04 & 87.85 & 87.53 & 87.33\\
    Difference & -0.41 & -0.24 & -0.21 & -0.05\\
    \hline
    \hline
    \end{tabular}
    \vspace{-5pt}
\end{table}

\subsubsection{Fidelity and Robustness under Non-i.i.d Settings}

In this section, we further evaluate the fidelity under non-i.i.d settings. All the experiments are conducted by training the VGG-16 models on the CIFAR10 dataset.

\vspace{0.3em}
\noindent\textbf{Fidelity under non-i.i.d settings.} The testing accuracy of the watermarked models and the models without the watermark is shown in Table \ref{table:noniid-fidelity}. The testing accuracy of the FedTracker models drops little compared with the baseline models without watermark. In non-i.i.d settings, the accuracy degradations are below $0.25\%$, which signifies that applying FedTracker does not significantly affect the model utility under non-i.i.d settings. 

\begin{figure}
    \centering
    \includegraphics[width=0.95\linewidth]{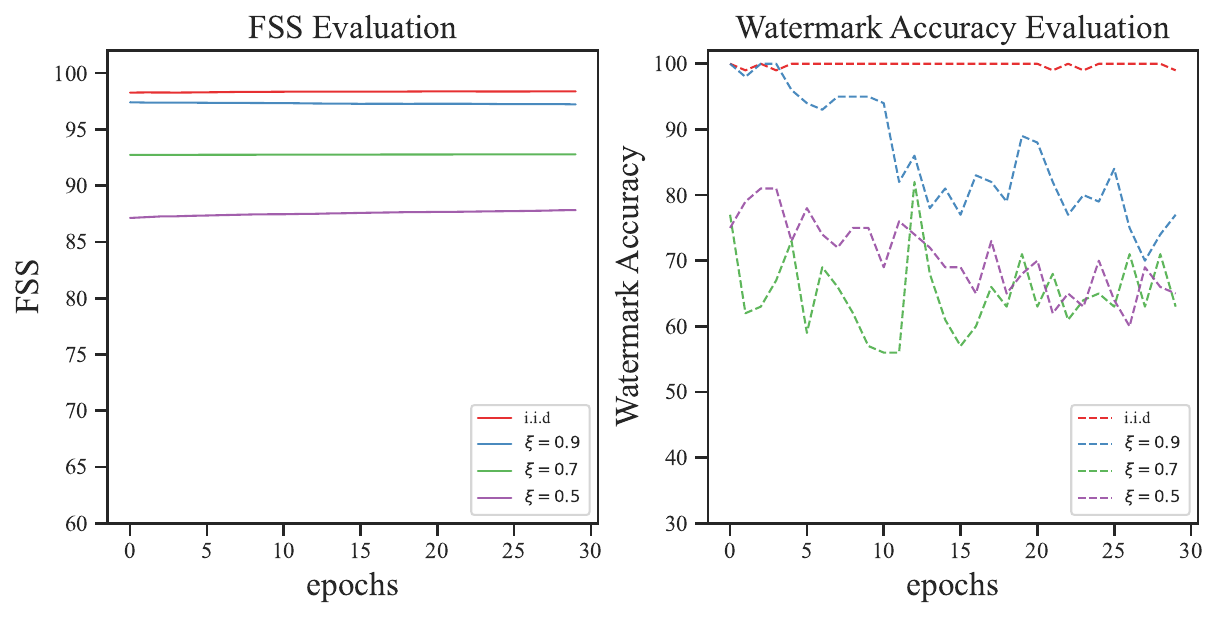}
    \caption{Watermark Accuracy and FSS Evaluation against Fine-tuning Attack under Non-i.i.d Settings.}
    \label{fig:finetuning-noniid}
    \vspace{-10pt}
\end{figure}

\vspace{0.3em}
\noindent\textbf{Robustness against fine-tuning attack under non-i.i.d settings.} Similar to the fine-tuning implementation in Section \ref{sec:robustness}, we fine-tune the Client's model with their own dataset. The difference is that in non-i.i.d settings, the Client's own data of different classes is unbalanced. The experimental results are shown in Fig. \ref{fig:finetuning-noniid}.

From the left part of Fig. \ref{fig:finetuning-noniid}, the fine-tuning attack still cannot affect the fingerprints embedded into the Batch Normalization layers. The FSS rarely changes during the fine-tuning, which indicates the great robustness of our method. From the right part of Fig. \ref{fig:finetuning-noniid}, we see that fine-tuning has an impact on the global watermark and the impact becomes larger as the variation increases. When $\xi=0.9$, the watermark accuracy drops to nearly $70\%$. And in the experiment of $\xi=0.5,0.7$, the lowest watermark accuracy is nearly $60\%$. It may be ascribed to the unbalanced data of the Client. Since we simulate the scenario that the Client itself fine-tunes the model on its own dataset, training on the unbalanced dataset leads the model to another local optimum. However, from the experiments, the watermark accuracy above $60\%$ is still enough for verification.

\vspace{0.3em}
\noindent\textbf{Robustness against Pruning Attack under Non-i.i.d Settings.} We also implement two different categories of pruning attacks. One is to prune the parameters of all the layers except the Batch Normalization layers, the other is to prune the parameters of the Batch Normalization layers. The results are shown in Table \ref{table:noniid-pruning1} and Table \ref{table:noniid-pruning2}. When pruning the parameters except those of the BN layers, both the watermark accuracy and the testing accuracy drop as the pruning rate increases. However, the watermark in the model trained under the non-i.i.d settings is more fragile than the model trained under the i.i.d setting. The watermark accuracy drops to $52\%$ when prunes $50\%$ parameters. Although the $52\%$ watermark accuracy can also verify the ownership, this phenomenon indicates that the non-i.i.d settings have an impact on the effectiveness to some extent.

From Table \ref{table:noniid-pruning2}, pruning the BN layers still cannot remove the watermark and fingerprint under non-i.i.d settings. Pruning BN layers significantly compromises the utility of the protected model. The testing accuracy drops more than $15\%$ which means the model has lost its value. Since it satisfies case~2 in Section \ref{sec:robustness}, the attack should not be considered a successful attack.

\begin{figure}[t]
    \centering
    \includegraphics[width=0.95\linewidth]{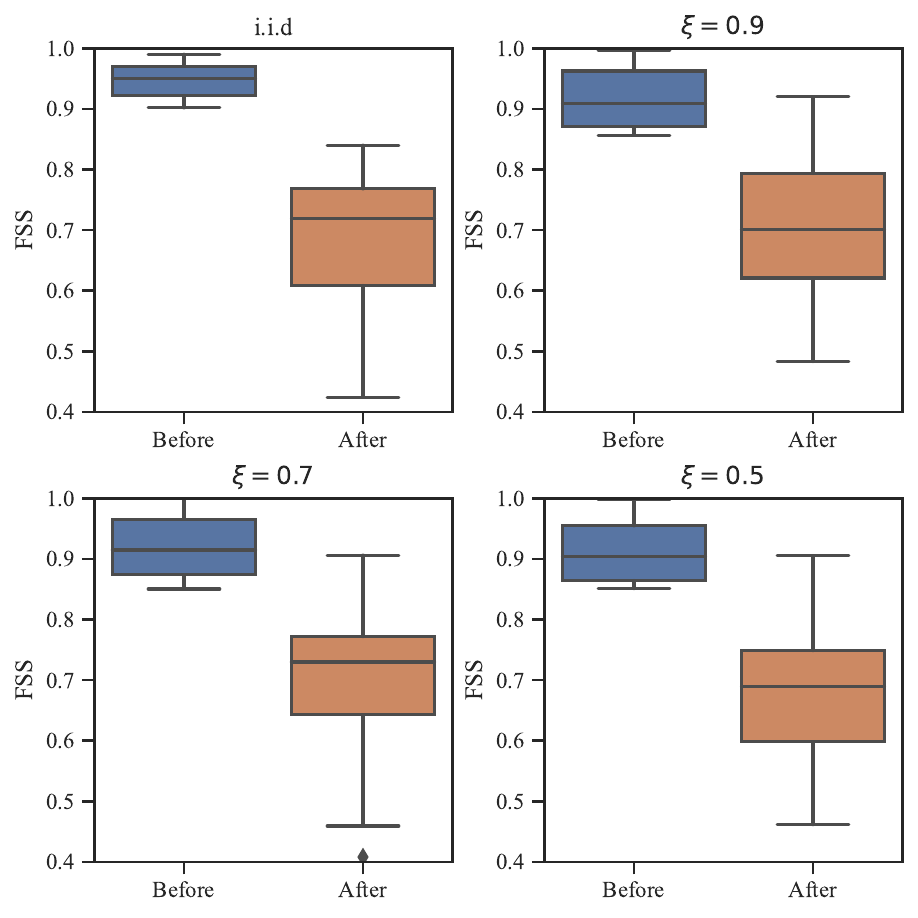}
    \caption{FSS Evaluation against Overwriting Attack under Non-i.i.d Settings.}
    \label{fig:overwrite-noniid}
    \vspace{-10pt}
\end{figure}

\begin{table*}[t]
    \centering
    \setstretch{1.1}
    \small
    \caption{Robustness against pruning attack which does not prune the BN layers.}
    \label{table:noniid-pruning1}
    \begin{tabular}{cc|ccccccccccc}
    \hline
    \hline
    & & \multicolumn{10}{c}{Pruning rate}\\
    Distribution & Metric & $0.05$ & $0.1$ & $0.15$ & $0.2$ & $0.25$ & $0.3$ & $0.35$ & $0.4$ & $0.45$ & $0.5$\\
    \hline
    \multirow{2}{*}{i.i.d} & WM acc & 100.00 & 100.00 & 100.00 & 100.00 & 100.00 & 99.00 & 99.00 & 98.00 & 94.00 & 83.00 \\
    & Testing acc & 89.69 & 89.7 & 89.59 & 89.58 & 89.40 & 89.16 & 88.45 & 87.83 & 86.35 & 84.42\\
    \hline
    \multirow{2}{*}{$\xi=0.9$} & WM acc & 100.00 & 100.00 & 100.00 & 100.00 & 100.00 & 100.00 & 99.00 & 99.00 & 94.00 & 79.00 \\
    & Testing acc & 85.46 & 85.38 & 85.37 & 85.63 & 85.66 & 85.88 & 85.63 & 85.17 & 84.34 & 82.32\\
    \hline
    \multirow{2}{*}{$\xi=0.7$} & WM acc & 100.00 & 100.00 & 100.00 & 100.00 & 98.00 & 98.00 & 96.00 & 85.00 & 77.00 & 69.00  \\
    & Testing acc & 85.82 & 85.79 & 85.89 & 85.79 & 85.70 & 85.30 & 85.16 & 84.47 & 82.38 & 81.06\\
    \hline
    \multirow{2}{*}{$\xi=0.5$} & WM acc & 84.00 & 84.00 & 84.00 & 81.00 & 77.00 & 78.00 & 78.00 & 77.00 & 71.00 & 52.00 \\
    & Testing acc & 83.69 & 83.63 & 83.31 & 83.60 & 83.50 & 83.54 & 83.57 & 83.82 & 83.38 & 81.11 \\
    \hline
    \hline
    \end{tabular}
    \vspace{-10pt}
\end{table*}

\begin{table*}[t]
    \centering
    \setstretch{1.1}
    \small
    \caption{Robustness against pruning attack which prunes the BN layers.}
    \label{table:noniid-pruning2}
    \begin{tabular}{cc|ccccccccccc}
    \hline
    \hline
    & & \multicolumn{10}{c}{Pruning rate}\\
    Distribution & Metric & $0.05$ & $0.1$ & $0.15$ & $0.2$ & $0.25$ & $0.3$ & $0.35$ & $0.4$ & $0.45$ & $0.5$\\
    \hline
    \multirow{3}{*}{i.i.d} & WM acc & 100.00 & 89.00 & 46.00 & 22.00 & 10.00 & 11.00 & 11.00 & 7.00 & 10.00 & 10.00\\
    & Testing acc & 89.69 & 84.41 & 72.94 & 56.41 & 32.88 & 27.06 & 25.33 & 17.47 & 11.47 & 12.75\\
    & FSS & 0.928 & 0.930 & 0.850 & 0.866 & 0.752 & 0.625 & 0.666 & 0.415 & 0.124 & 0.025 \\
    \hline
    \multirow{3}{*}{$\xi=0.9$} & WM acc & 100.00 & 64.00 & 14.00 & 10.00 & 10.00 & 10.00 & 10.00 & 10.00 & 10.00 & 10.00 \\
    & Testing acc & 85.46 & 82.09 & 68.08 & 49.75 & 39.58 & 33.59 & 25.02 & 11.81 & 10.83 & 10.01\\
    & FSS & 0.974 & 0.974 & 0.972 & 0.968 & 0.961 & 0.955 & 0.941 & 0.922 & 0.946 & 0.946\\
    \hline
    \multirow{3}{*}{$\xi=0.7$} & WM acc & 100.00 & 56.00 & 11.00 & 10.00 & 10.00 & 10.00 & 10.00 & 10.00 & 10.00 & 10.00 \\
    & Testing acc & 85.82 & 80.00 & 65.17 & 51.66 & 42.40 & 33.36 & 13.24 & 10.33 & 10.12 & 11.59\\
    & FSS & 0.927 & 0.927 & 0.924 & 0.926 & 0.923 & 0.909 & 0.904 & 0.887 & 0.848 & 0.876\\
    \hline
    \multirow{3}{*}{$\xi=0.5$} & WM acc & 84.00 & 47.00 & 12.00 & 12.00 & 11.00 & 11.00 & 10.00 & 10.00 & 10.00 & 10.00\\
    & Testing acc & 83.69 & 77.60 & 68.75 & 48.77 & 41.55 & 32.05 & 23.65 & 22.65 & 13.44 & 11.25\\
    & FSS & 0.870 & 0.871 & 0.869 & 0.850 & 0.850 & 0.832 & 0.819 & 0.817 & 0.772 & 0.766 \\
    \hline
    \hline
    \end{tabular}
    \vspace{-10pt}
\end{table*}

\vspace{0.3em}
\noindent\textbf{Robustness against Overwriting Attack under Non-i.i.d Settings.} We also conduct the overwriting attack on the models trained under non-i.i.d settings to evaluate if the non-i.i.d settings make the fingerprint fragile. The results are shown in Fig. \ref{fig:overwrite-noniid}. We still achieve $100\%$ traceability rate which means FedTracker can still disclose the model leaker against overwriting attack. Moreover, in all four experiments, the decrease in FSS is almost the same. The median values of FSS are all reduced to around $0.7$. This indicates that the robustness against the overwriting attack of the model trained under non-i.i.d settings does not change.

To sum up, under non-i.i.d settings, the fidelity and robustness of the FedTracker watermark and fingerprints are not significantly influenced. This proves that FedTracker is widely applicable to a variety of non-i.i.d scenarios.

\subsection{Comparing FSS and Hamming Distance}
\label{sec:comparing}

In FedTracker, we propose to use the Fingerprint Similarity Score (FSS) to measure the similarity between different fingerprints. However, the existing watermarking scheme uses Hamming distance (HD) as a metric. In this section, we present a detailed comparison of FSS and HD.

\begin{table}[t]
  \centering
  \setstretch{1.05}
  \caption{Traceability rate (\%) using FSS and HD.}
  \label{table:hd}
  \small
  \begin{tabular}{cc|cccc}
  \hline
  \hline
  & & \multicolumn{4}{c}{Number of Clients} \\
  Model & Metric & 10 & 20 & 30 & 50\\
  \hline
  \multirow[c]{2}{*}{CNN-4} & FSS & 100.0 & 100.0 & 100.0 & 100.0\\
  & HD & 100.0 & 100.0 & 100.0 & 100.0\\
  \hline
  \multirow[c]{2}{*}{AlexNet} & FSS & 100.0 & 100.0 & 100.0 & 100.0\\
  & HD & 100.0 & 100.0 & 100.0 & 100.0\\
  \hline
  \multirow[c]{2}{*}{VGG-16} & FSS & 100.0 & 100.0 & 100.0 & 100.0\\
  & HD & 100.0 & \emph{90.0} & 100.0 & 100.0\\
  \hline
  \multirow[c]{2}{*}{ResNet-18} & FSS & 100.0 & 100.0 & 100.0 & 100.0\\
  & HD & 100.0 & 100.0 & \emph{23.3} & 100.0\\
  \hline
  \hline
  \end{tabular}
  \vspace{-10pt}
\end{table}

When verifying the ownership of the deep learning model which is trained by only one party, the model owner only needs to compare the watermark extracted from the suspicious model with one single watermark. In that case, HD is enough to measure the similarity. However, in FedTracker, we need to compare the local fingerprint with each Client's fingerprint. HD rounds the value to an integer and thus loses some information. For example, the distance $0.7$ and the distance $1.3$ are both rounded to $1$, making the two fingerprints indistinguishable.

The experimental results of using FSS and HD as the metric are shown in Table \ref{table:hd}. From Table \ref{table:hd}, we can find that in most cases, HD works as well as FSS. But in the cases that the VGG-16 trained with 20 Clients and ResNet-18 trained with 30 Clients, using HD fails to disclose the identity of the model leaker while using FSS is always effective. The results demonstrate that utilizing FSS can better discriminate different fingerprints.


\section{Conclusion}
\label{sec:conclusion}

In this paper, we propose FedTracker, the first FL model protection framework to simultaneously furnish ownership verification and traceability. Global watermark and local fingerprint mechanisms are utilized to verify the ownership of the FL model and trace the model back to the leaker respectively. We accordingly design a CL-based watermark embedding mechanism to improve fidelity and a novel metric named Fingerprint Similarity Score to measure the similarity of fingerprints. Evaluations of various tasks and settings show that FedTracker is effective in ownership verification, traceability, fidelity, and robustness.


%

\ifCLASSOPTIONcompsoc
  \section*{Acknowledgments}
\else
  \section*{Acknowledgment}
\fi

This research is supported by the National Key Research and Development Program of China (2020AAA0107705) and the National Natural Science Foundation of China (62072395, U20A20178).


\bibliographystyle{IEEEtran}
\bibliography{IEEEabrv, reference}

\end{document}